\pdfoutput=1
\documentclass[11pt,a4paper,tightenlines,eqsecnum,floats,aps,amsmath,amssymb,nofootinbib,prd,showpacs,superscriptaddress]{revtex4-1}

\usepackage{epsfig}
\usepackage{setspace}
\usepackage{amsmath,amssymb,amsfonts,amsthm}
\usepackage{graphicx}
\usepackage{multirow}
\usepackage{xfrac}
\usepackage[justification=centering]{caption}
\usepackage{subcaption}
\usepackage{changepage}
\usepackage[yyyymmdd]{datetime}
\usepackage{graphicx}
\usepackage{float}
\usepackage{verbatim}
\usepackage{lmodern}
\usepackage{pgfplots}
\usepackage{colortbl}
\usepackage{lipsum}
\usepackage{stmaryrd}

\newcommand{\be}{\begin{equation}}
\newcommand{\ee}{\end{equation}}

\newcommand{\rmd}{{\rm d}}
\newcommand{\mn}{{\mu \nu}}
\newcommand{\nn}{\nonumber \\}

\newdateformat{monthyeardate}{%
	\monthname[\THEMONTH] \THEDAY, \THEYEAR}

\makeatletter

\begin{document}
\begin{flushleft}
  
\end{flushleft}
\vspace{1cm}
\title{Gravitational Wave Bursts from Cosmic String Cusps and Pseudocusps}

\author{Matthew J. Stott} 
\email{matthew.stott@kcl.ac.uk} 
	\affiliation{Theoretical Particle Physics and Cosmology Group, Department of Physics, \\King's College London, University of London, Strand, London, \\WC2R 2LS, United Kingdom} 
\author{Thomas Elghozi}		\email{thomas.elghozi@kcl.ac.uk}
	\affiliation{Theoretical Particle Physics and Cosmology Group, Department of Physics, \\King's College London, University of London, Strand, London, \\WC2R 2LS, United Kingdom} 
\author{Mairi Sakellariadou}\email{mairi.sakellariadou@kcl.ac.uk}
	\affiliation{Theoretical Particle Physics and Cosmology Group, Department of Physics, \\King's College London, University of London, Strand, London, \\WC2R 2LS, United Kingdom} 
	\affiliation{Perimeter Institute for Theoretical Physics, Waterloo, Ontario N2L 2Y5, Canada}

\date{\monthyeardate\today}

\thanks{}
\begin{abstract}
We study the relative contribution of cusps and pseudocusps, on cosmic (super)strings, to the emitted bursts of gravitational waves. The gravitational wave emission in the vicinity of highly relativistic points on the string follows, for a high enough frequency, a logarithmic decrease. The slope has been analytically found to be $\sfrac{-4}{3}$ for points reaching exactly the speed of light in the limit $c=1$. We investigate the variations of this high frequency behaviour with respect to the velocity of the points considered, for strings formed through a numerical simulation, and we then compute numerically the gravitational waves emitted. We find that for string points moving with velocities as far as $10^{-3}$ from the theoretical (relativistic) limit $c=1$, gravitational wave emission follows a behaviour consistent with that of cusps, effectively increasing the number of cusps on a string. Indeed, depending on the velocity threshold chosen for such behaviour, we show the emitting part of the string worldsheet is enhanced by a factor ${\cal O}(10^3)$ with respect to the emission of cusps only.
\end{abstract}

\maketitle

\newpage
\section{Introduction}
Applying Grand Unified Theories (GUTs) in the context of the early Universe, one concludes that as the Universe expands and its temperature drops, it undergoes a series of phase transitions followed by spontaneously broken symmetries, which may leave behind topological defects as false vacuum remnants~\cite{Kibble,Hind_Kibble,Vilenkin_shellard,ms-cs07,Vachaspati:2015cma}. Cosmic strings are one-dimensional such topological defects, generically formed after a period of hybrid inflation~\cite{Jeannerot:2003qv}; they can lead to a variety of observational consequences. In the context of string theoretic models, brane interactions can lead to fundamental strings, one-dimensional Dirichlet branes, and their bound states, which are collectively referred to as cosmic superstrings~\cite{Copeland:2003bj,PolchRevis,Sakellariadou:2008ie}. These objects, which could be formed for instance at the end of brane inflation~\cite{Burgess:2001fx,Sarangi:2002yt,Lizarraga:2016hpd}, are the quantum analogs of the field theoretic one-dimensional topological defects and can play a similar cosmological role.

It has been shown using analytical and numerical means that the networks of cosmic (super)strings reach a scaling regime, in which the overall energy density of such network remains constant throughout cosmic evolution. The presence of sub-horizon strings (loops)~\cite{Ringeval:2005kr,Lorenz:2010sm,Blanco-Pillado:2013qja}, in addition to super-horizon ones (sometimes referred to as infinite strings), does not alter this feature, nor does the formation of Y-junctions~\cite{TWW,Rajantie:2007hp,Copeland:2006if,Urrestilla:2007yw,Sakellariadou:2008ay,Copeland:2007nv,PACPS}, which mainly occur on superstrings. Indeed, while collisions of cosmic strings lead the exchange of partners with probability equal to $1$, cosmic superstrings have an intercommutation probability several orders of magnitude smaller and can therefore entangle and form junctions~\cite{PolchProb,cmp,JJP}. Such a smaller intercommutation probability will effect the string network evolution~\cite{Sakellariadou:2004wq,Avgoustidis:2005nv}.

The main phenomenological consequence of a string network is the emission of Gravitational Waves (GWs)~\cite{Vachaspati:1984gt,Sakellariadou:1990ne,Damour2000,Damour2001,Damour2004,Brandenberger:2008ni,Olmez:2010bi,Binetruy:2010cc,Regimbau:2011bm,Aasi:2013vna}, generating bursts as well as a stochastic background of gravity waves. Indeed, several features such as cusps (points temporarily reaching the speed of light $c=1$), kinks (discontinuities of the tangent vector created by intercommutation processes) and junctions, produce high frequency GW Bursts (GWBs). Given the large mass per unit length a string could possess, cusps, receiving sizeable Lorentz boosts, offer rich possibilities for detectable signals from gravitational wave emissions. In particular, it has been shown~\cite{Damour2000} that the spectrum of bursts from cusps presents a logarithmic decrease following a slope of $\sfrac{-4}{3}$ for high frequencies. In order to estimate the signal one could possibly expect to receive on Earth or in its neighbourhood, it is necessary to relate the occurrence of cusps to such a model and its network parameters, for instance the string tension and the inter-string distance. Complimentary to this, it is also of importance to understand in greater detail which points should be considered for such emission processes.

We aim in this paper to give an analysis of the high frequency slope dependence of the GW spectrum with respect to the velocities of the emitting points on the string, using numerical simulations. Indeed, to our knowledge, this has only been done though an analytical means and only for points defined as exact cusps, that is, points reaching exactly the velocity of light $c=1$. Equivalently, one can define a cusp as the equality between the left- and right-movers' derivatives, under such decomposition of the string position vector. Still, it has been found~\cite{ENS2014} that a significant number of important emitters would be ignored if only cusps were to be considered. So-called \emph{pseudo}cusps are points reaching highly relativistic velocities below the speed of light, related to a point of close approach of the left- and right-movers' derivatives. As these points are also thought to take part in emission of GWBs, we intend to quantify both their importance and to refine their definitions with respect to GWs.

In the following, we first recall in Section~\ref{sec:setup}, the setup we choose in order to study this problem and the analytical results already known so far. We also review the Kibble-Turok approach~\cite{Kibble:1982cb,Turok:1984cn}
 to string trajectories along with the important stages of Damour and Vilenkin analytical study of GWB~\cite{Damour2000}. In Section~\ref{sec:numerics}, we give the details of the approach and numerical methods we used for our study while in Section~\ref{sec:results} we analyse and discuss the findings we obtained. In particular, we show that a variety of points around cusps but also points independent from cusps (namely, not in a cusp's vicinity), should be taken into account when considering the GW emissions from cusps on cosmic (super)strings. Finally in Section~\ref{sec:conclusions} we conclude our findings.

\section{String dynamics and gravitational waveforms from cusps and pseudocusps} \label{sec:setup}

Cosmic (super)strings have been shown to appear in two main forms, sub-horizon strings (loops), whose size is of the order of the coherence length scale, and super-horizon (infinite) ones whose length is greater than the horizon scale. Sub-horizon string can self-interact, or two super-horizon strings can intersect, forming chopping smaller string loops. In the presence of semi-local string interactions or in the context of string theory, strings can entangle to form bound states limited by Y-junctions.

If we are to consider a cosmic string whose length scale is sufficiently large as compared to its thickness, any long range interactions between the portions of the string can be disregarded enabling the string to be accurately modelled as a one dimensional object. Under this approximation the string sweeps out a two-dimensional surface known as the string worldsheet. We model the dynamics of a string using the Nambu-Goto approximation, that is, in the limit that their thickness goes to zero, following the Nambu-Goto action
\begin{equation} \label{eq:nambu}
	S = -\mu \int {\rm d} \tau \,\rmd \sigma \,\sqrt{-\gamma} ~,
\end{equation}
where $\mu$ is the string tension, $\tau, \sigma$ are the worldsheet coordinates (timelike and spacelike, respectively), and $\gamma$ is the induced metric on the string worldsheet. We will then choose to impose the conformal (Virasoro) gauge conditions,
\begin{align}
	(\dot{X}_\mu)^2 + (X_\mu')^2 = 0 \label{eq:Vira1} \\
	\dot{X}^\mu \,X_\mu' = 0 \label{eq:Vira2} ~,
\end{align}
along with the temporal gauge constraint $\tau = t \equiv X^0$. These conditions lead to the following string equation of motion
\begin{equation}
	\bf{X}'' - {\bf\ddot{X}} = {\bf 0}~.
\end{equation}
To solve this, we use the lightlike coordinates $\sigma_\pm \equiv \sigma \pm t$, and decompose the position vector $\bf{X} (\sigma, \tau)$ into left- and right-movers, ${\bf X}_+ \left( \sigma_+ \right)$, ${\bf X}_- \left( \sigma_- \right)$, as
\be \label{eq:x-world}
	{\bf X} \left( t , \sigma \right) \equiv \frac{1}{2} \left[ {\bf X}_+ \left( \sigma_+ \right) + {\bf X}_- \left( \sigma_- \right) \right]
\ee
hence
\begin{align}
	{\bf X}' \left( t , \sigma \right) &= \frac{1}{2} \left[ \dot{\bf X}_+ \left( \sigma_+ \right) + \dot{\bf X}_- \left( \sigma_- \right) \right] \\
	\dot {\bf X} \left( t , \sigma \right) &= \frac{1}{2} \left[ \dot{\bf X}_+ \left( \sigma_+ \right) - \dot{\bf X}_- \left( \sigma_- \right) \right] ~,
\end{align}
where the over-dot denotes a derivative with respect to the time coordinate $t$, while the dash denotes a derivative with respect to the space coordinate $\sigma$ or with the only null coordinate $\sigma_\pm$. Recall that the Virasoro conditions in this gauge are equivalent to $|\dot{\bf X}_+| = 1 = |\dot{\bf X}_-|$ where the worldsheet coordinates give $X^0(\tau,\sigma)$ such that $X_+^0 = \sigma_+$, $X_-^0=\sigma_-$. It has been shown~\cite{ENS2014} that loops and strings stretched between fixed junctions with heavy strings have a periodic or pseudo-periodic motion, that is, the curves described by the vectors $\dot{\bf X}_+$ and $\dot{\bf X}_-$ are closed on the unit sphere.

The phenomena we are interested in are cusps and \emph{pseudo}cusps, that is, the highly relativistic points on the string. Cusps are defined as points whose velocity reaches $c=1$, which is equivalent to an exact crossing of the two curves $\dot{\bf X}_+$ and $-\dot{\bf X}_-$ on the unit sphere, namely
\be \label{eq:convention}
	\dot{\bf X}_+ (\sigma_+^{\rm (c)}) = -\dot{\bf X}_- (\sigma_-^{\rm (c)}) ~,
\end{equation}
the superscript $\rm (c)$ denoting the cusp's coordinates. Numerically, this exact equality is difficult to achieve, therefore one can allow for some leeway. In addition, \emph{pseudo}cusps have been defined~\cite{ENS2014} as points reaching a highly relativistic velocity but corresponding to an exact velocity of $c=1$, for instance any velocity $10^{-6}$ to $10^{-3}$ below $c=1$. Recall that this definition was partly arbitrary and related to the numerical accuracy of the simulation used in the work presented in Ref.~\cite{ENS2014}. The nature of such a definition is something we wish to clarify in this paper.

An imperative point to examine is that such velocities could be reached in the neighbourhood of cusps, since the velocity function $|\dot {\bf X} (\sigma,t)|$ is continuous in both $t$ and $\sigma$. As such, one would have $\sigma_+^{\rm (pc)} = \sigma_+^{\rm (c)} + \delta \sigma_+$ and $\sigma_-^{\rm (pc)} = \sigma_-^{\rm (c)} + \epsilon \,\delta \sigma_-$ where $\epsilon \equiv \pm 1$ and $|\delta \sigma_\pm| \ll \sigma_{\rm max}$, $\sigma_{\rm max}$ being either the periodicity of $\dot{\bf X}_\pm$ or some coherence scale of $\dot{\bf X}_\pm$ on the unit sphere.\footnote{For $\delta \sigma_+ = \delta \sigma_- = \delta \sigma > 0$, if $\epsilon = 1$, the pseudocusp appears at the same time as the cusp but further on along the string, with $\sigma^{\rm (pc)} = \sigma^{\rm (c)} + \delta \sigma$ and $t^{\rm (pc)} = t^{\rm (c)}$; on the contrary, if $\epsilon = -1$, it is at the point of the cusp but slightly after it occurred, that is, $\sigma^{\rm (pc)} = \sigma^{\rm (c)}$ and $t^{\rm (pc)} = t^{\rm (c)} + \delta \sigma$.}. In addition, a further extension to the definition of such points could be pseudocusps appearing \emph{per se}, in a region whose velocity will never achieve $c=1$ exactly. This phenomena would be apparent in the neighbourhood of a close approach between the two curves on the unit sphere, without any direct crossing. The inclusion of points defined in this way will greatly increase the number of points one might choose to look at as opposed to enforcing the limit $c=1$.
\begin{figure}[t]
	\includegraphics[width=10cm, keepaspectratio, trim= 225 275 120 120, clip]{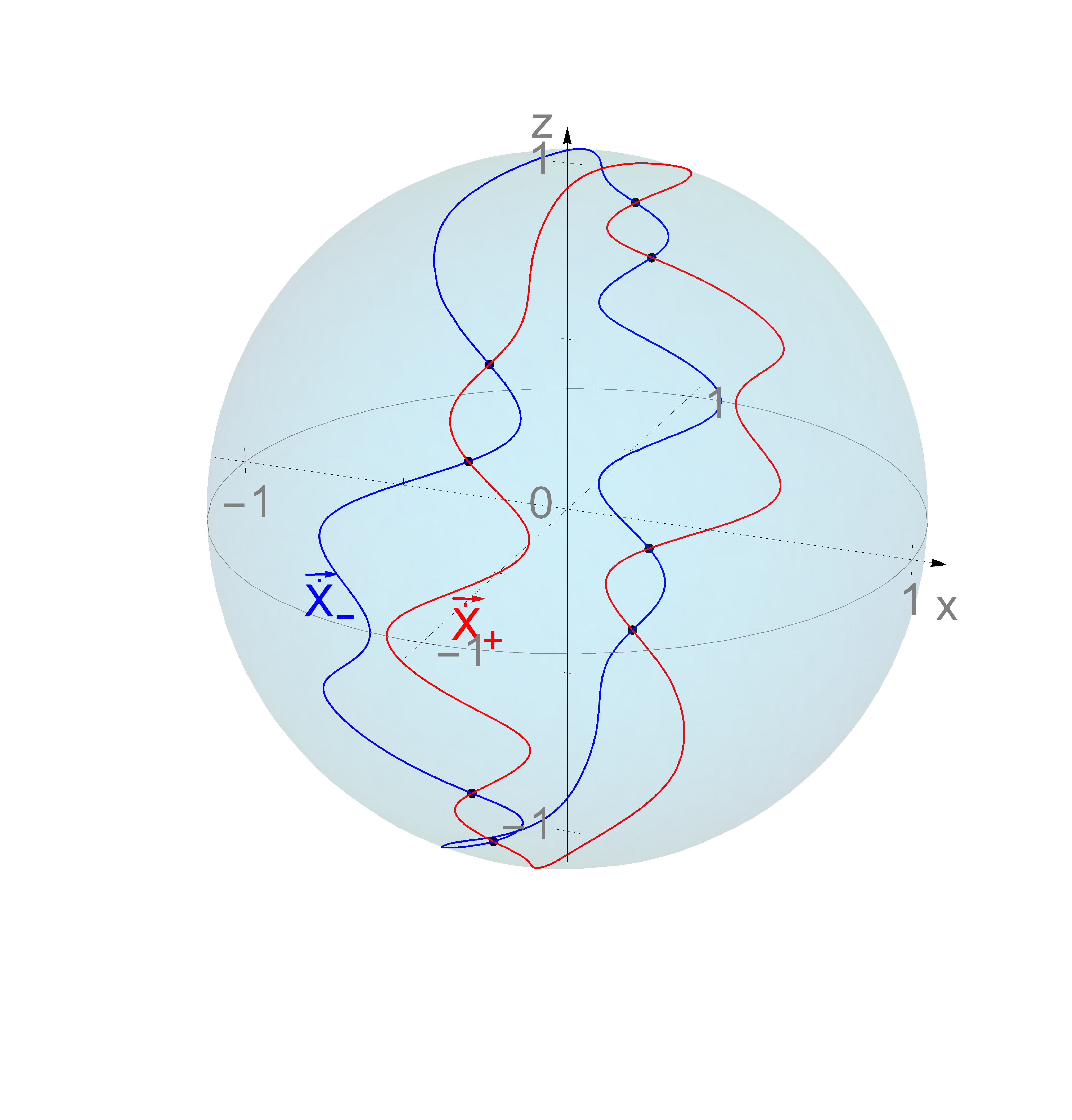}
	\caption{Schematic representation of the $\dot{\bf X}_+$ (in red) and $\dot{\bf X}_-$ (in blue) curves on the unit sphere, with black points highlighting the cusps configurations, that is, the crossings of these curves.}
	\label{fig:unitexample}
\end{figure}

Cosmic (super)strings are strong emitters of various radiations (depending on the model one has in mind), in particular GWs, since their energy density can be very high and their motion relativistic. Every point of every string contained in the cosmic network they form, thus emits GWs, generating a stochastic background. We will be directing our interest to bursts of GWs, and in particular those produced by cusps and pseudocusps, since they have been shown to be dominating the high frequency end of the spectrum, with a logarithmic decrease presenting a slope $\sfrac{-4}{3}$. It has been estimated~\cite{ENS2014}, using numerical simulations and analytical work, that pseudocusps should be in significant proportions with respect to cusps. In this study therefore we wish to address the ability to estimate the importance of GWBs from points classified as pseudocusps using our previous definitions.

\subsection{Pseudocusps}

As we mentioned already, cusps, which are points of the string reaching temporarily $c=1$, are maxima of the two-dimensional velocity function $\left| {\bf \dot X} \right| (\sigma, t)$. Pseudocusps, which are defined as points reaching a highly relativistic velocity (strictly) below $c=1$, belong to one of the two classes, defined as follows. Firstly, points located in the neighbourhood of a cusp: indeed, because the velocity function $\left| {\bf \dot X} \right| (\sigma, t)$ is continuous in both variables, the immediate vicinity of the maximum $\left| {\bf \dot X} \right| (\sigma^{\rm (c)}, t^{\rm (c)}) = 1$ satisfies $\left| {\bf \dot X} \right| (\sigma^{\rm (pc)}, t^{\rm (pc)}) \lesssim 1$, where one observes an infinitesimal deviation in the spatial and temporal directions, as in $\sigma^{\rm (pc)} = \sigma^{\rm (c)} + \delta \sigma$ and $t^{\rm (pc)} = t^{\rm (c)} + \epsilon \,\delta t$, where $\epsilon \equiv \pm 1$ and $|\delta \sigma|, \,|\delta t| \ll \sigma_{\rm max}$. Secondly, pseudocusps can also be independent maxima of such velocity function: points at which the derivatives of the left- and right-movers' ($\dot{\bf X}_+$ and $\dot{\bf X}_-$) are approaching very close to each other, enough for such points to be highly relativistic, but not getting exactly equal, before moving apart of each other. Points in the latter class are not located in the vicinity of a cusp, they are hence called independent pseudocusps. Their own neighbourhood is necessarily highly relativistic as well. For such points, we define $\sigma_\pm^{\rm (pc)} = \sigma^{\rm (pc)} \pm t^{\rm (pc)}$ to be the (null) coordinates at which the velocity is a local maximum. Said differently, $(\sigma_+^{\rm (pc)}, \sigma_-^{\rm (pc)})$ is where these two vectors $\dot{\bf X}_\pm$ are locally the closest, and we denote by $\theta^{\rm (pc)}$ the angle between them
\be
	\theta^{\rm (pc)} \equiv \arccos \left( \dot{\bf X}_+ (\sigma_+^{\rm (pc)}) \cdot \dot{\bf X}_- (\sigma_-^{\rm (pc)}) \right) ~.
\ee
We also define~\cite{ENS2014},
\begin{align}
	l^\mu \equiv \dot X^\mu (\sigma^{\rm (pc)}, t^{\rm (pc)}) = \frac{1}{2} \left( {\dot X}_+^\mu (\sigma^{\rm (pc)}_+) - {\dot X}_-^\mu (\sigma^{\rm (pc)}_-) \right) \\
	\delta^\mu \equiv \frac{1}{2} \left( {\dot X}_+^\mu (\sigma^{\rm (pc)}_+) + {\dot X}_-^\mu (\sigma^{\rm (pc)}_-) \right) ~,
\end{align}
the half-sum and the half-difference between the left- and right-movers' (4-vectors) derivatives, computed at the point of interest, here a pseudocusp. Note that despite its appearance, we call $l^\mu$ the half-\emph{sum}, recalling the vectors we are interested in are ${\dot X}_+^\mu$ and $- {\dot X}_-^\mu$.\footnote{Still, because of symmetries of the form $\sigma_\pm \rightarrow 2 {\cal L} - \sigma_\pm$, one can generally study the $+{\dot X}_\pm^\mu$ curves on the unit sphere, provided a careful computation of the event coordinates, e.g. $\sigma_\pm^{\rm (c)}$. See Fig.~\ref{fig:periodcon} and the discussion around it.}

The 4-vector $l^\mu$ is defined as the 4-velocity at the point of interest and we recall that it is a null vector in the case of a cusp. In the case of pseudocusps, the time-component $l^0$ is also equal to $1$, but the norm of the 3-velocity of the string at that point $(\sigma^{\rm (pc)}_+, \sigma^{\rm (pc)}_-)$ equals
\be \label{eq:PCveltheta}
	|l^i| = \sqrt{\frac{1 + \cos(\theta_c)}{2}} \approx 1 - \sfrac{\theta_c^2}{8}~.
\ee
However, $\delta^\mu$ is spacelike, with $\delta^0 = 0$ in the time gauge, and
\be
	|\delta^i| = \sqrt{\frac{1 - \cos(\theta_c)}{2}} \approx \sfrac{\theta_c}{2}~.
\ee
The angle $\theta_c$ can be thought of as measuring the \emph{softness} of a relativistic part of the string. The larger it is, the smaller the velocity and the softer the pseudocusp; in the limit $\theta_c = 0$, the event recovers a cusp such that the velocity reaches $c=1$.

Note that in Ref.~\cite{ENS2014}, only independent pseudocusps (local maxima whose velocity is below $c=1$) were being looked for, the pseudocusps lying in the vicinity of cusps being ignored. It was found, under some hypothesis that will be reviewed here, that there exists a significant number of pseudocusps (about half as many) in comparison with the number of cusps.

\subsection{Gravitational wave radiation}
 
Our numerical simulation closely follows the assumptions and analytical results by Damour and Vilenkin in Ref.~\cite{Damour2000,Damour2001,Damour2004} as well as work in Ref.~\cite{Olmez:2010bi} in order to consider the waveforms of both points defined as cusps and pseudocusps. We start by considering the linearised metric perturbation, $g_\mn = \eta_\mn + h_\mn$, with $\eta_\mn$ defined as the flat metric and $|h_\mn| \ll 1$, where we choose to set the harmonic gauge, $\partial^\nu h_{\mu\nu} = 0$. Considering the traceless part, gravitational waves from a localised source can be calculated using the linearised Einstein equations
\begin{equation}
	\Box h_{\mu\nu} = -16 \pi G \,T_{\mu \nu} ~,
\end{equation}
which, for the far-field approximation can be expressed up to first order in $\sfrac{1}{r}$ as
\begin{equation}
	h_{\mu\nu} \simeq \frac{4G}{r} \sum_\omega e^{-i\omega(t-r)} \,T_{\mu\nu}~.
\end{equation}
Varying the Nambu-Goto action given in Eq.~(\ref{eq:nambu}) with respect to the metric yields the string energy-momentum tensor for any position vector $x^\lambda$
\begin{equation}
	T^{\mu\nu}(x^\lambda) = \mu \int \rmd \tau \rmd \sigma \left( \dot{X}^\mu \dot{X}^\nu - X'^{\mu} X'^{\nu} \right) \delta^{(4)}(x^\lambda -X^\lambda(\tau,\sigma)) ~,
\end{equation}
where the dependences of $X^\mu$, $X'^\mu$ and ${\dot X}^\mu$ in $\sigma$ and $t$ have been hidden and $\delta^{(4)}$ is the four dimensional Dirac delta function. In the Fourier domain, for any point $k^\lambda$ in momentum space, one obtains
\begin{equation} \label{eq:T-fourier}
	T^{\mu\nu}(k^\lambda) = \frac\mu{\sfrac{\cal L}{2}} \int_{\Sigma} \rmd \tau \rmd \sigma \left( \dot{X}^\mu \dot{X}^\nu - X'^\mu X'^\nu \right) e^{-ik\cdot X} ~,
\end{equation}
where $\cal L$ is the period in $\sigma_{\pm}$ and $\Sigma$ is the compact string worldsheet $(\sigma,t) \in [0, \sfrac{\cal L}{2}]^2$. Note the 4-vector contraction $k \cdot X \equiv k^\lambda X_\lambda$. Using the left- and right-movers decomposition and the worldsheet coordinates given in Eq.~(\ref{eq:x-world}), we obtain
\begin{equation} \label{eq:Tmn}
	T^{\mu\nu}(k^\lambda) = \frac\mu{\cal L} \int_{\bar \Sigma} \rmd \sigma_+ \rmd \sigma_- \left( \dot{X}^{(\mu}_+ \,\dot{X}^{\nu)}_- \right) e^{-\frac{i}{2}(k\cdot X_+ + k\cdot X_-)} ~,
\end{equation}
where we introduced the symmetrisation over the indices $\mu$ and $\nu$ and again choose not to explicitly write the dependence of $X_\pm^\mu$ and ${\dot X}_\pm^\mu$ in $\sigma_\pm$. Here, $\bar \Sigma$ is the compact worldsheet using the light-like coordinates $(\sigma_+,\sigma_-) \in [0, {\cal L}]^2$. Using a complete factorisation of the integrand, we arrive at the final form for the stress-energy tensor
\begin{equation}
	T^{\mu\nu}(k^\lambda) = \frac{\mu}{\cal L} \,I^{(\mu}_+ I^{\nu)}_- ~, \\
\end{equation}	
where the factors appearing in the stress-energy tensor which will be of interest in our study are
\begin{equation}
	\qquad I^\mu_{\pm} \equiv \int^{\cal L}_0 \rmd \sigma_{\pm} \,\dot{X}^\mu_{\pm} \,e^{-\frac{i}{2}k\cdot{X_\pm}} ~. \label{eq:I_tensor}
\end{equation}
Due to the fact that the periodicity $\cal L$ is either the invariant length of the string $L$, or it is of the order a few times this length at most~\cite{ENS2014}, and since in our simulation will set ${\cal L} = L$ (without any loss of generality), we shall assume this equality from this point onwards.

A final remark on the integration interval: as noted in Ref.~\cite{Damour2001}, the majority of the integral above comes from the small interval around the coordinates $\sigma_\pm = 0$, that is, around the position of a cusp or a pseudocusp. One can therefore formally expand the limits of the integral to the full length of the string when studying the high frequency behaviour. In our code, this is where the precision can be optimised (at most $200,000$ points over the interval $[0, L]$). Similarly, the number of integrations performed is set by the number of frequency points considered with in the region $[f_u, f_l]$ (see below, Eq.~(\ref{eq:flimits})).

\subsection{Waveform frequency dependence and gauge term considerations}

The main steps derived in Ref.~\cite{Damour2001} which detail the frequency dependence of the waveform from (exact) cusps have been reproduced in this section. Let us start by recalling that the direction of maximal emission is along the velocity direction at the cusp, such that $k^\lambda = \omega \,\dot{X}^\lambda (\sigma^{\rm (c)}, t^{\rm (c)}) \equiv \omega \,l^\lambda$, with $\omega$ the GW frequency.\footnote{Recall the ${\rm (c)}$ superscript denoting the position of the cusp} This is the case of interest to us from now on.

The string's left- and right-movers, $X_{\pm}^\mu (\sigma_\pm)$, can be expanded around the point of intersection on the unit sphere, and truncated at third order (in order to keep the first non-null term in the final expression, as we will see below). After a shift of worldsheet coordinates such that $\sigma_{\pm}^{\rm (c)} = 0$ and of spacetime coordinates such that $X_{\pm}^\mu (0) = 0$, one obtains
\begin{align}
	X_{\pm}^\mu (\sigma_{\pm}) = \pm \,l^\mu \,\sigma_{\pm} + \frac{1}{2} \ddot{X}^\mu_{\pm} (0) \,\sigma_{\pm}^2 + \frac{1}{6} \dddot{X}_{\pm}^\mu (0) \,\sigma^3_{\pm} \label{eq:loc1} \\
	\dot{X}^\mu_{\pm}(\sigma_{\pm}) = \pm \,l^\mu + \ddot{X}^\mu_{\pm} (0) \,\sigma_{\pm} + \frac{1}{2} \dddot{X}^\mu_\pm (0) \,\sigma^2_{\pm} ~, \label{eq:loc2}
\end{align}
where $l^\mu \equiv {\dot X}^\mu (\sigma^{\rm (c)}, t^{\rm (c)}) = \pm {\dot X}_\pm^\mu (\sigma_\pm^{\rm (c)})$ since we are considering a cusp here.

Using the Virasoro conditions $\dot X^2_\pm (\sigma_\pm) = 0$, which one can differentiate to obtain $\dot{X}_{\pm} \cdot \ddot{X}_{\pm} = 0$ and $\dot{X}_{\pm} \cdot \dddot{X}_{\pm} + \ddot{X}_{\pm}^2 = 0$ at any $\sigma_\pm$, one gets, at the cusp (only), $l \cdot \ddot{X}_{\pm} (0) = 0$ and $l \cdot \dddot{X}_{\pm} (0) = -\ddot{X}_{\pm}^2 (0)$. One can use these to express the quantity which enters in the phase of Eq.~(\ref{eq:I_tensor}) as
\begin{equation}
	k \cdot {X_\pm} (\sigma_\pm) = \omega \,l_\mu X^\mu_{\pm}(\sigma_{\pm}) = -\frac{1}{6} \omega \,(\ddot{X}_{\pm}^\mu)^2 \,\sigma^3_{\pm} ~.
\end{equation}
In addition, the three leading terms in the integrand of Eq.~(\ref{eq:Tmn}), of the form $l^{(\mu} l^{\nu)} + l^{(\mu} \ddot{X}^{\nu)}_- (0) \,\sigma_- + \ddot{X}^{(\mu}_+ (0) l^{\nu)} \,\sigma_+$, are purely gauge terms and as such must be gauged away~\cite{Damour2001}. The integrals $I^\mu_\pm$ then become
\begin{equation}
	I_{\pm}^\mu = \ddot{X}_{\pm}^\mu \int^L_0 \rmd \sigma_{\pm} \,\sigma_{\pm} \,e^{\frac{i}{12} \omega \,\ddot{X}^2_\pm \,\sigma_\pm^3} ~,
\end{equation}
where $\ddot{X}_{\pm}^\mu$ is the leading order (gauge removed) term. Computing these integrals yields
\begin{equation}
	I_{\pm}^\mu = \frac{8\pi i}{\Gamma(\frac{1}{3})} \frac{\ddot{X}^\mu_{\pm}}{|\ddot{X}_{\pm}|^{\frac{4}{3}}} \,\frac{1}{\omega^{\frac{2}{3}}}~.
\end{equation}
Simply replacing $\omega = 2 \pi f$ reveals the frequency power law for the quantities $I^\mu_{\pm}$ and for the gravitational waves energy-momentum tensor
\begin{align}
\label{eq:solvedtensor}
	I^\mu_{\pm} &= i A^\mu_{\pm} \,\omega^{-\frac{2}{3}} \\
	T^\mn (k^\lambda = \omega \,l^\lambda) &= - \frac{\mu}{L} A^{(\mu}_+ A^{\nu)}_- \,|\omega|^{-\frac{4}{3}} 
\nonumber\\
	&= - \frac{64 \pi^2}{\Gamma (\sfrac{1}{3})^2} \frac{\mu}{L} \frac{{\ddot X}^{(\mu}_+ \,{\ddot X}^{\nu)}_-}{\left[ \left| {\ddot X}^\alpha_+ \right| \,\left| {\ddot X}^\beta_- \right| \right]^{\sfrac{4}{3}}} \;\omega^{\sfrac{-4}{3}}
\end{align}
where $A_\pm^\mu \equiv \frac{8\pi}{\Gamma(\frac{1}{3})} \frac{\ddot{X}^\mu_{\pm}}{|\ddot{X}_{\pm}|^{\frac{4}{3}}}$ are functions of the movers' second derivative $\ddot{X}^\mu_{\pm}$. The important feature one should extract from the above expressions is the final form of the frequency dependence.

As we have shown the waveform for cusps has been found to have a frequency dependence in which the stress-energy tensor follows a $f^{\sfrac{-4}{3}}$ power law ($f^{\sfrac{-1}{3}}$ in the logarithmic Fourier representation)~\cite{Damour2001}. In order to attempt to reproduce this dependence in our code, where $c=1$ is never really achieved because of numerical inaccuracy, we must ensure that we include only corresponding physical components of the GWB. We are considering the point of strongest emission, $\sigma_\pm^{(s)}$, in the direction of strongest emission $k^\lambda \propto l^\lambda \equiv \dot{X}^\lambda (\sigma^{(s)}, t^{(s)})$. Recall we noticed that, under a local Taylor expansion, the leading order terms are pure gauge and do not represent any physical properties for GWB considerations. We will take that into account, in the decomposition of the quantities we use in our numerical code. Using the expansion and applying the Virasoro conditions, we now have
\begin{align}
	I^\mu_{\pm} &= \int^{l}_{0} \rmd \sigma_{\pm} \left( \dot{X}^\mu_{\pm} (\sigma_\pm) \mp l^\mu \right) e^{-\frac{i}{2}k \cdot X_{\pm} (\sigma_\pm)} \nn
	&= \int^{l}_{0} \rmd \sigma_{\pm} \left( \dot{X}^\mu_{\pm} (\sigma_\pm) \mp \frac{1}{2} \left( {\dot X}_+^\mu (\sigma^{(s)}_+) - {\dot X}_-^\mu (\sigma^{(s)}_-) \right) \right) e^{-\frac{i}{2} \omega \,l \cdot X_{\pm} (\sigma_\pm)} ~, \label{eq:gauged}
\end{align}
where the $\mp$ sign is due to cusps occurring at $\dot{X}^\mu_+ = - \dot{X}^\mu_-$ in our conventions. This definition of the integrals allows us to also consider points which do not fully satisfy the conditions set by cusps, that is a full realisation of Eq.~(\ref{eq:convention}), thus allowing for the study of pseudocusps in a similar manner.
Given the now correctly determined quantities $I^\mu_{\pm}$, we will use the following scalar measure in order to evaluate GWB amplitudes
\begin{align}
	T &= \sqrt{T^\mn T^*_\mn - \frac{1}{2} \left| T^\mu_{\;\mu} T^{*\nu}_{\;\nu} \right|} ~, \nonumber \\
	&= \frac{\mu}{L} \,\sqrt{\left(I^{(\mu}_+I^{\nu)}_-\right) \left( I^*_{+(\mu } I^*_{- \nu)} \right) - \frac{1}{2} \left| \left(I^{(\mu}_+I_{- \mu)}\right) \left(I^{*(\nu }_+I^*_{- \nu)}\right) \right|} ~, \label{eq:scalar2}\\
	\mathcal{T} &= \frac{L}{\mu} \,T = \sqrt{\left(I^{(\mu}_+I^{\nu)}_-\right) \left( I^*_{+(\mu } I^*_{- \nu)} \right) - \frac{1}{2} \left| \left(I^{(\mu}_+I_{- \mu)}\right) \left(I^{*(\nu }_+I^*_{- \nu)}\right) \right|} ~, \label{eq:scalar}
\end{align}
where $\mathcal{T}$ is the quantity of interest we shall use in our analysis in order to appraise the waveform's slope dependence.

\section{Numerical simulation configuration and Methodology} \label{sec:numerics}

\begin{figure}[t]
	\includegraphics[width=11cm, keepaspectratio, trim= 150 120 20 10, clip]{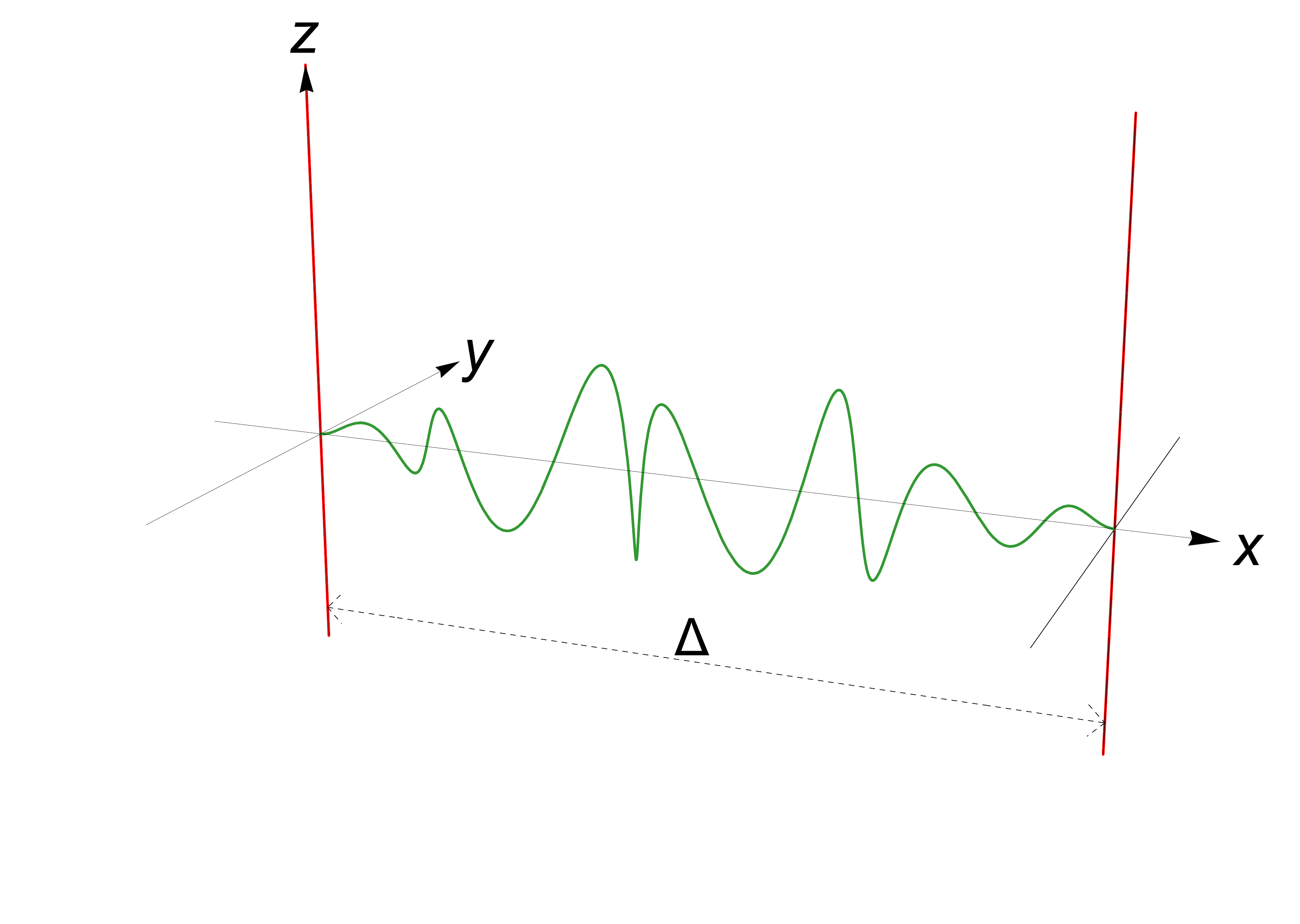}
	\caption{Visualisation of string configuration used in our numerical simulations depicting a light string stretched between two junctions with heavier fixed strings.}
	\label{fig:stringconfiguration}
\end{figure}

Our study comprises of developing a numerical code in order to evaluate the previously derived form of the stress-energy tensor coming from Eq.~(\ref{eq:scalar}) based on, in part, the simulation used in Ref.~\cite{ENS2014}. Let us first recall some of its properties before detailing the new computations. Our initial simulation studies the non-interacting movement of light strings stretched between two junctions with heavier, fixed strings as displayed in Fig.~\ref{fig:stringconfiguration}. The end points of our simulated string can thus only move in the $z$-direction. As the movement of the string is assumed periodic, without any loss of generality~\cite{ENS2014}, its initial position is described by a sum of $\bar n_i \leq n$ Fourier modes, with amplitude $c^i_k$ (for the cosines) and $s^i_k$ (for the sines), as in
\begin{align}
	{\bf X} (\sigma,0) = \frac{\sigma}{\sigma_m} \,{\bf \Delta} &+ \left[ \sum_{k=1}^{{\bar n}_x} c^x_k (h_m) \cos \left( \frac{2 \pi k \,\sigma}{\sigma_m} \right) + s^x_k (h_m) \sin \left( \frac{2 \pi k \,\sigma}{\sigma_m} \right) \right] {\bf e}_x \nn
	&+ \left[ \sum_{k=1}^{{\bar n}_y} c^y_k (h_m) \cos \left( \frac{2 \pi k \,\sigma}{\sigma_m} \right) + s^y_k (h_m) \sin \left( \frac{2 \pi k \,\sigma}{\sigma_m} \right) \right] {\bf e}_y \nn
	&+ \left[ \sum_{k=1}^{{\bar n}_z} c^z_k (h_m) \cos \left( \frac{2 \pi k \,\sigma}{\sigma_m} \right) \right] {\bf e}_z ~,
\end{align}
where ${\bf \Delta} = (\Delta, \,0, \,0)$, ${\bar n}_i$ are random integers uniformly drawn from $\llbracket 1,n \rrbracket$ and all $c^i_k (h_m)$ and $s^i_k (h_m)$ yield random real numbers uniformly drawn from $[ -h_m, \,h_m ]$. $n$ and $h_m$ are input parameters of the simulation; they set up the oscillatory behaviour of the string, fixing a limit to the highest frequency and amplitudes reached in its Fourier decomposition.

Note that while these amplitudes are all drawn in $[-h_m,\,h_m]$, with a uniform distribution, there is a bias: indeed, too large amplitudes can sometimes lead to velocities temporarily above $c=1$ (for instance if a large amplitude is drawn for several similar high frequencies). Still, the evolution equation implies that, if at $t=0$ the velocity is well-behaved and below $c=1$, it will remain so during the whole period. The wrongly-behaved strings are dismissed, therefore distorting {\it a posteriori} the uniform draw within the amplitude interval. It is also important to remark that our choice of uniform distribution may affect the probability of cusps and pseudocusps, as it may favour high frequencies in comparison to, for instance, a Gaussian distribution; this is not a problem given this should not distort the proportion of cusps to pseudocusps.

The initial velocity is obtained in a similar way. Indeed, one can remark first that the norm of the velocity vector ${\bf \dot X} (\sigma,0)$ can be computed using the tangent vector ${\bf X}' (\sigma,0)$ and the Virasoro condition $({\bf \dot X})^2 + ({\bf X'})^2 = 1$ (which is thus automatically satisfied). To fix its direction, we rotate it within the orthogonal plane to the tangent vector ${\bf X}'$. The other Virasoro condition ${\bf \dot X} \cdot {\bf X'} = 0$ is then also satisfied by such a construction. To assure both continuity and periodicity, the rotation angle $\alpha (\sigma)$ is also given by a Fourier decomposition: a number of amplitudes are uniformly drawn from the interval $[\alpha_m, \alpha_m]$ where one gets $\alpha (\sigma) = \sum_{k=1}^{{\bar n}_\alpha} s^\alpha_k (\alpha_m) \sin \left( \frac{2 \pi k \,\sigma}{\sigma_m} \right)$. The boundary conditions, which impose a direction of the velocity at the end points, are satisfied before and after such a rotation. One has now obtained ${\bf X}' (\sigma,0)$ and ${\bf \dot X} (\sigma,0)$, $\forall \sigma$, and the equation of motion leads to the decomposition $\dot{\bf X}_\pm (\sigma) = {\bf X}' (\sigma,0) \pm {\bf \dot X} (\sigma,0)$, $\forall \sigma \in [0,2\sigma_m]$. This yields the complete (non-interacting) evolution of the string over a period of time. Note that we also check that various constraints, such as the Virasoro conditions, are indeed well satisfied within the whole $(t,\sigma) \in [0, \sigma_m]^2$ plane.

\begin{figure}[t]
	\includegraphics[width=10cm,keepaspectratio]{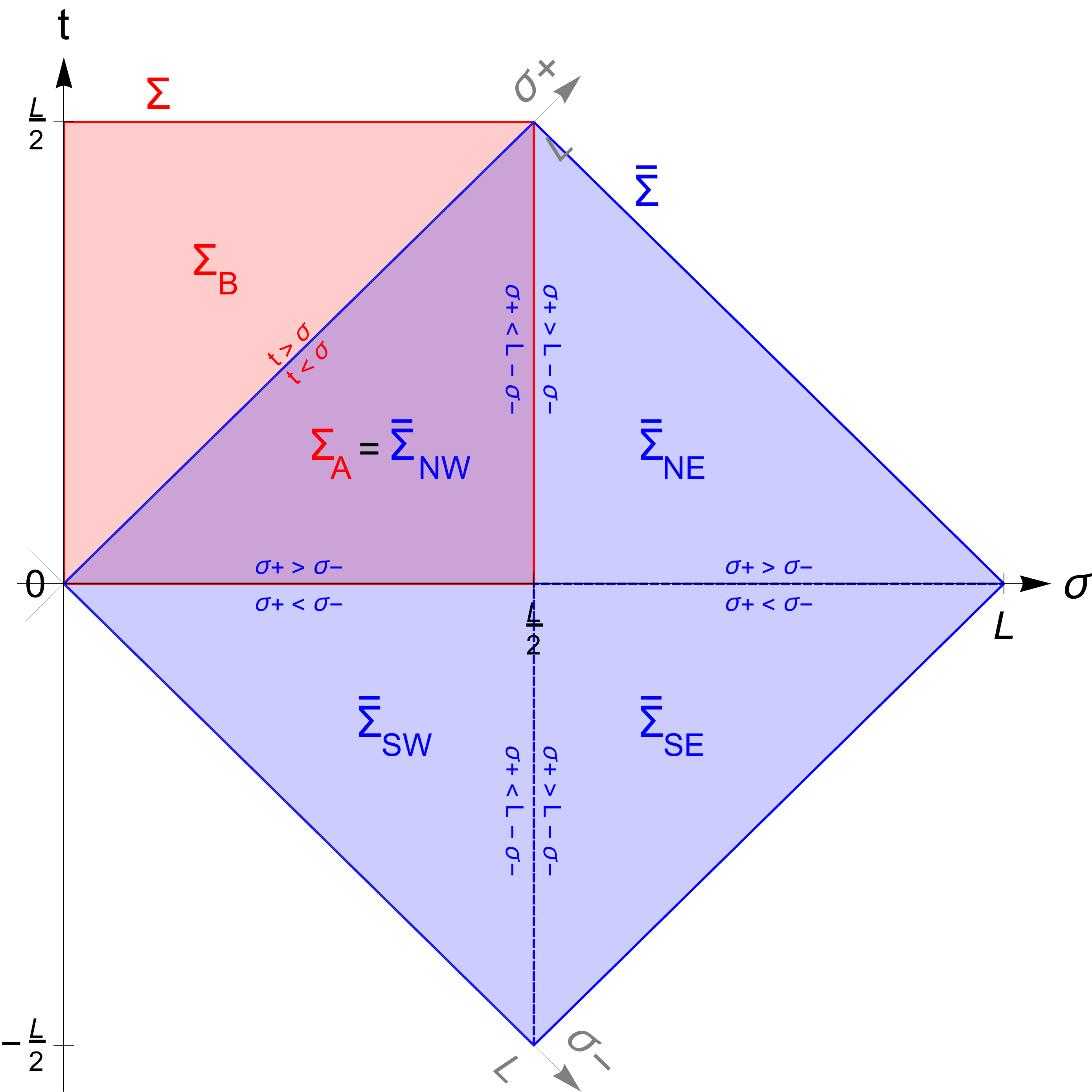}
	\caption{String worldsheets using space and time coordinates ($\Sigma$ in red) and null coordinates ($\bar \Sigma$ in blue). Even though $\bar \Sigma$ is twice as large as $\Sigma$, both contain the same information. Indeed, $\Sigma_{\rm A} = {\bar \Sigma}_{\rm NW}$ and ${\bar \Sigma}_{\rm SW}$ are the symmetric of ${\bar \Sigma}_{\rm NE}$ and ${\bar \Sigma}_{\rm SE}$ (respectively), under the symmetry $\sigma \rightarrow L - \sigma$; also, the symmetric of $\Sigma_{\rm B}$ (using again $\sigma \rightarrow L - \sigma$) is equivalent to ${\bar \Sigma}_{\rm SW}$ due to the $L$-periodicity of $\bar \Sigma$.}
	\label{fig:periodcon}
\end{figure}

It is important to note that while the null coordinate worldsheet $\bar \Sigma \equiv \left\{ (\sigma_+, \sigma_-) \in [0, L]^2 \right\}$ is twice (and not four times) as large as its space and time coordinates counterpart $\Sigma \equiv \left\{ (\sigma, t) \in [0, \sfrac{L}{2}]^2 \right\}$, there exist symmetry properties, coming from the boundary conditions of our numerical silumation, implying that all the information within $\Sigma$ appears exactly twice in $\bar \Sigma$. Indeed, recall first that while the string's motion is $\sfrac{L}{2}$-periodic in time, the curves $\bf{ \dot X}_{\pm}$ are $L$-periodic. In addition, the string itself has been drawn such that its velocity vector is symmetric under the transformation $\dot{X}^\mu(\sigma,t) = \dot{X}^\mu(L-\sigma,t)$. As shown in Fig.~\ref{fig:periodcon}, let us divide the $\Sigma$ worldsheet in two parts according to $t \lessgtr \sigma$, as in $\Sigma_{\rm A} \equiv \left\{ (\sigma, t) \in [0, \sfrac{L}{2}]^2, t < \sigma \right\}$ and $\Sigma_{\rm B} \equiv \left\{ (\sigma, t) \in [0, \sfrac{L}{2}]^2, t > \sigma \right\}$; similarly, let us divide $\bar \Sigma$ in four parts according to $\sigma_+ \lessgtr \sigma_-$ and $\sigma_+ \lessgtr L - \sigma_-$, and denote them ${\bar \Sigma}_{\rm NW} = \Sigma_{\rm A}$, ${\bar \Sigma}_{\rm NE}$, ${\bar \Sigma}_{\rm SE}$ and ${\bar \Sigma}_{\rm SW}$. Using the aforementioned symmetries, one can note that ${\bar \Sigma}_{\rm NW}$ and ${\bar \Sigma}_{\rm SW}$ are respectively the symmetric of ${\bar \Sigma}_{\rm NE}$ and ${\bar \Sigma}_{\rm SE}$ (using $\sigma \rightarrow L - \sigma$). In addition, the symmetric of $\Sigma_{\rm B}$ (using $\sigma \rightarrow L - \sigma$) is the equivalent of ${\bar \Sigma}_{\rm SW}$ (using the periodicity in $\sigma_+$). Therefore, $\Sigma$ is half of $\bar \Sigma$, whose study yields twice each distinct point (such as cusps) appearing in the study of $\Sigma$.

The total number of cusps forming on a simulated string is found by analysing the $\dot{\bf X}_\pm$ curves on the unit sphere and looking for direct crossings between these vectors. The velocity is then computed and checked as to whether it reaches the required limit of $c = 1$ within the numerical uncertainties of our code, which are generally $\lesssim 10^{-6}$. The identified independent pseudocusps are the other points of local maximum with highly relativistic velocities. Arbitrarily {\it a priori} we followed from previous work in Ref.~\cite{ENS2014} was considering ``highly relativistic'' to be any velocity above $0.999 \,c$. Finally, it is checked that pseudocusps correspond to configurations with a very small separation between the two curves on the unit sphere. The angle $\theta^{\rm (pc)}$ between the two vectors $\dot{\bf X}_+$ and $- \dot{\bf X}_-$ is computed and its minimum found (within the grid approximation). This generally lies between $10^{-1}$ and $10^{-3}$, in agreement with Eq.~(\ref{eq:PCveltheta}) and our definition of pseudocusps with respect to their velocity. To be clear following the work in Ref.~\cite{ENS2014} cusps are accordingly defined as points corresponding to direct crossings spanning the velocity range $1 c \rightarrow 0.999999 c$. Independent pseudocusps are points corresponding to near crossings spanning the velocity range $0.999999 c \rightarrow 0.999c $. Given we are interested in the behaviour of points in the neighbourhood of the velocity limit $c=1$, which could potentially deviate by several orders of magnitude, we refer to point velocities on the string using the scale $\log(1-v)$, ergo $-\infty < \log(1-v) \leq -6$ for cusps and $-6 < \log(1-v) \leq -3$ for independent pseudocusps. The significance and validity for these limits is something we shall explore in this paper. 

Finally, its worth noting that even though our analysis is performed within a specific setup, our qualitative results remain valid in the more realistic string configurations, most importantly in the direct environment of a cusp or pseudocusp as well as the relative proportion of such events. Based on these simulations, that is, starting with a sample of strings with cusps and independent pseudocusps occurring during a period of non-interacting movement, we computed the energy-momentum tensor of the GWs emitted. As the high frequency end of the spectrum is insensitive to the low velocity portions of the string, one can chose to integrate over the whole string, as we did, or over a small region around the point of interest. In addition, the accuracy of our results highly depends on the accuracy of the initial code used to produce our simulated strings and the cuspy (cusps and pseudocusps) configurations, which is of order $10^{-6}$.

Before we look at the statistical implications of our numerical simulations, we will briefly cover any approximations or considerations we take over the full process for the study of points on a single string. The frequency window we limit ourselves to in our numerical computations is defined by the following modification to the relation in Eq.~(\ref{eq:solvedtensor}), yielding
\be \label{eq:flimits}
	T^\mn (f) \propto A^{(\mu}_+ A^{\nu)}_- \,|f|^{-\frac{4}{3}} \;\Theta(f_u-f) \Theta(f-f_l) ~,
\ee
where the step functions $\Theta$ serve as a cut in the upper and lower frequency bounds, respectively $f_u \simeq 10^4~{\rm Hz}$ and $f_l \simeq 10~{\rm Hz}$. The purpose of the lower frequency limit will be to disregard a frequency plateau at the low end of the spectrum as our aim is to focus on the high frequency burst emissions coming from highly relativistic ($\log(1-v) < -3$) parts of the string. Regular (non-relativistic, with $\log(1-v) < -1$) parts of the string serve to generate a stochastic background of GWs, emitted in no preferred direction and with an amplitude almost independent from $f$. Similarly, our upper limit removes a region of computational uncertainty at large frequencies (roughly above $10^4~{\rm Hz}$), where our code was found to be unreliable as it yields very noisy, unrealistic outputs. We believe this is due to the numerical inaccuracies coming from ${\dot X}^\mu_\pm$ or higher order derivatives, which appear in the exponential part of the integrand of $I_\pm^\mu$ which blown up as $\omega$ becomes increasingly large. This leaves a (conservative) several-orders-of-magnitude frequency window in which highly relativistic points would emit GWBs and where we fully trust our simulation's results, namely $\omega \in [10,10^4]$. It is in this region we expect to match the analytical computations leading to the $\sfrac{-4}{3}$ linear decrease in the log power spectrum as shown by Damour and Vilenkin through our numerical simulations.

It is important to note that while variations in acceptable bounds between each relativistic region could allow for larger windows, there is no relation between the upper and lower bounds for each point of each string. Therefore, we have chosen safer, more stringent limits to ensure our automated treatment is considering the correct (stable, high end) frequencies for the study of $\mathcal{T}$, reducing the window on each end by two points on a logarithmic scale, from $\omega \in [10,10^4]$ to $\omega \in [16, 6400]~{\rm Hz}$.\footnote{More accurately, the chosen interval is $[10^{1.2}, 10^{3.8}] = [15.9,6310]~{\rm Hz}$.} This procedure leaves us with a $2.6$-order-of-magnitude investigation window, that is, for the frequency $f = \sfrac{\omega}{2 \pi}$ in Eq.~(\ref{eq:gauged}), an interval $f \in [2.52, 1000]$ Hz, where we are considering the waveform slope dependance for different highly relativistic points.

\begin{figure}[t]
	\begin{adjustwidth}{-5em}{-5em}
	\begin{center}
	\begin{subfigure}{9cm}
		\includegraphics[width=0.99\linewidth]{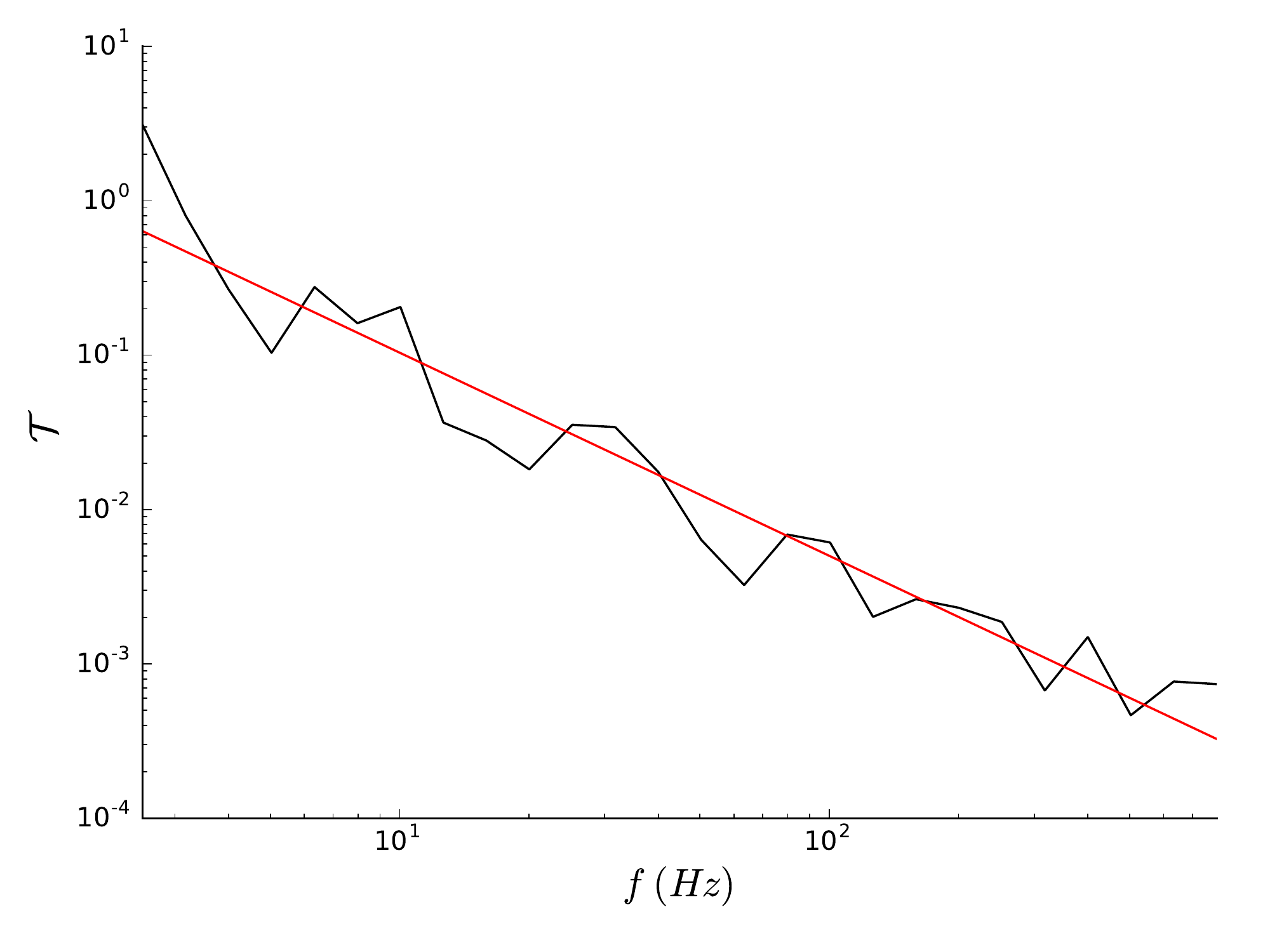}
		\caption{26 frequency points yielding a calculated slope of -1.3158}
		\label{fig7:a}
		\vspace{4ex}
	\end{subfigure}
	\begin{subfigure}{9cm}
		\includegraphics[width=0.99\linewidth]{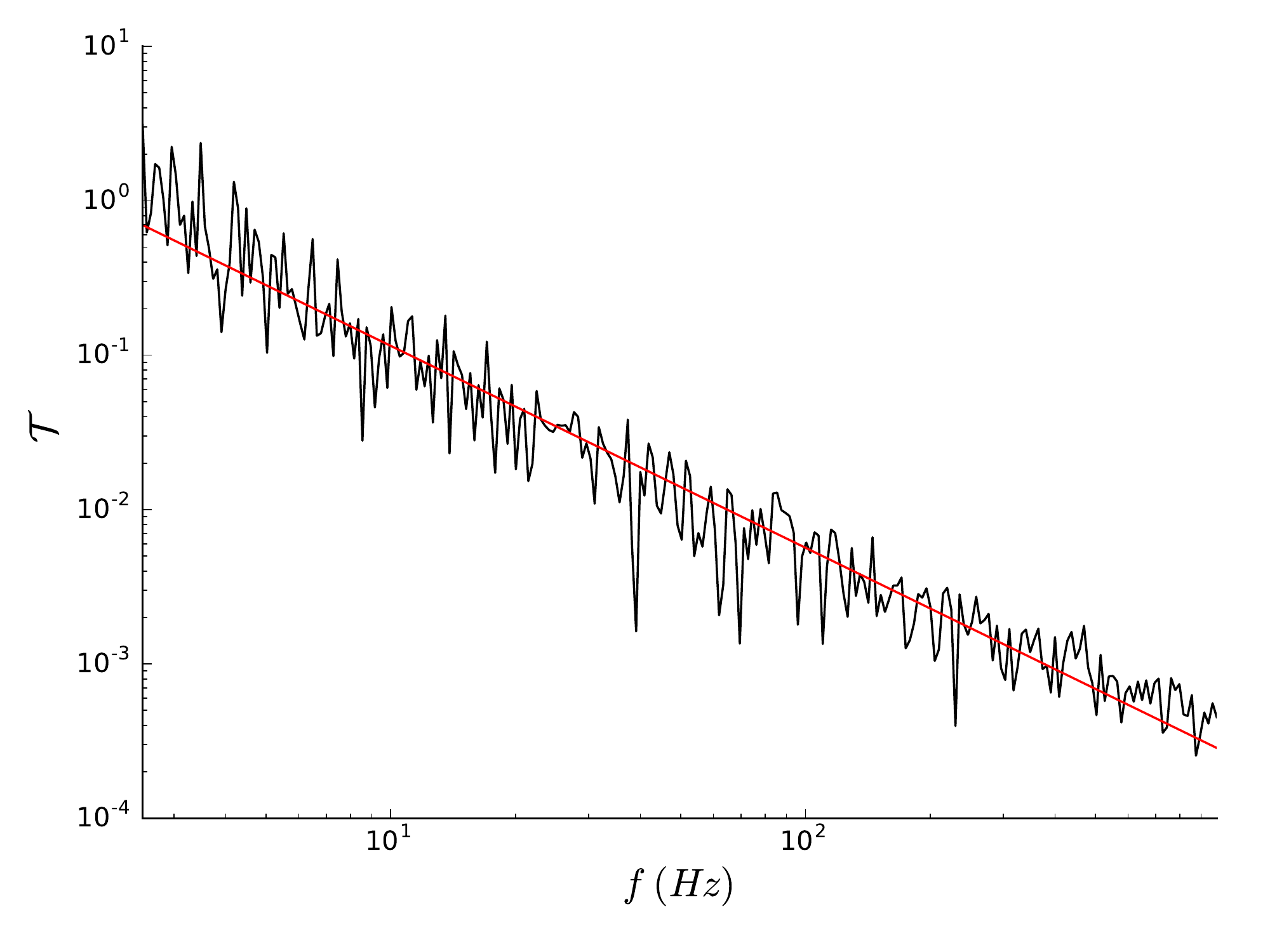}
		\caption{260 frequency points yielding a calculated slope of -1.3076}
		\label{fig7:b}
		\vspace{4ex}
	\end{subfigure}
	\begin{subfigure}{9cm}
		\includegraphics[width=0.99\linewidth]{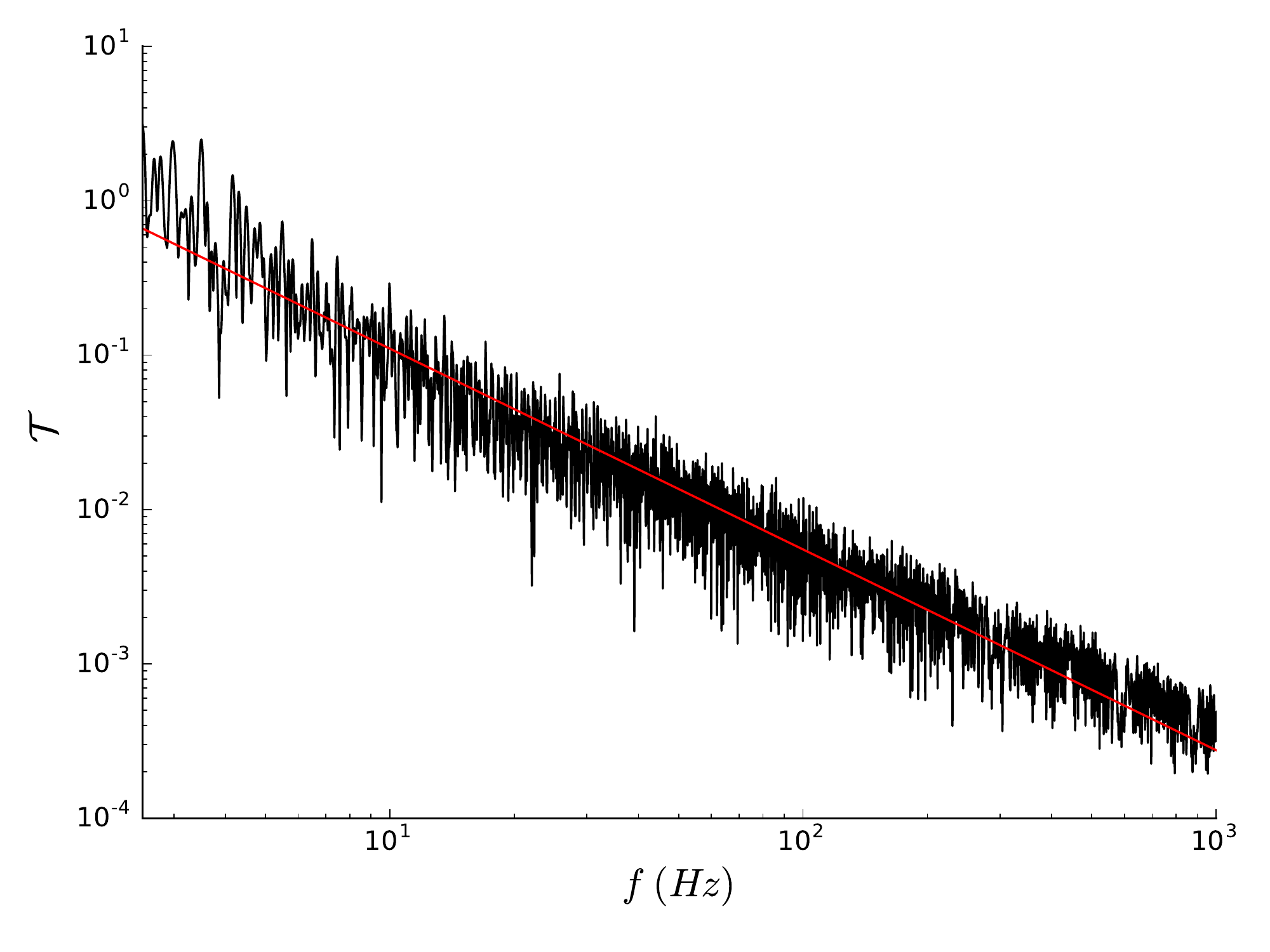}
		\caption{2600 frequency points yielding a calculated slope of -1.3005}
		\label{fig7:c}
	\end{subfigure}
	\end{center}
	\end{adjustwidth}
\caption{Calculated slope for a single cusp using a different number of frequency points in the frequency interval $[2.52, 1000]~{\rm Hz}$ defined in Section~\ref{sec:numerics}. We find that the slope converges quickly enough that significantly optimising computational time with a lower number of points is an acceptable approximation allowing for an analysis of a much larger sample of points.}
	\label{fig:numfreq}
\end{figure}

Figure~\ref{fig:numfreq} details the convergence of the calculated slope for a single cusp as the potential number of frequency points we could consider, within our defined window, is varied over three orders of magnitude. As the main focus of this paper is to gain a statistical understanding of the overall behaviour of a range of highly relativistic points, we limit ourselves to the approximation of the slope consisting of the lower number of frequency points, as we find this is sufficient to fully encode the slopes behaviour at these points, whilst allowing for significantly decreased computational time. This in turn greatly increases the number of both strings and points we are able to consider. Our code is therefore limited to perform integrations for 26 logarithmically equidistant frequency points. The increasingly present numerical noise we observe in each plot in Fig.~\ref{fig:numfreq} comes from the sensitivity in the exponential in Eq.~(\ref{eq:gauged}) to the numerical inaccuracies of our code manifest from the accuracy of decomposition in the strings coordinates.\footnote{Figure~\ref{fig:string1plots} in App.~\ref{sec:codeout} presents the output for every cusp identified on a single string when taking into account these considerations, which serves to confirm our methodology holds valid across a large number of points.}

\section{Numerical simulation results and discussion} \label{sec:results}

Our code has considered a total of 119 simulated strings in which we identified 3234 highly relativistic points corresponding to either direct or near crossings above the thresholds given in Section~\ref{sec:numerics}. Of these, a total of 2123 points were classified as cusps and 1111 points were classified as independent pseudocusps. We have also investigated a further 7088 points corresponding to velocities below the lower velocity threshold for pseudocusps by a single order of magnitude ($-3 < log(1-v) \leq -2$) in order to fully decipher the transitional behaviour and apparent limits between the highly relativistic and non-relativistic regimes. Indeed, we seek to identify a velocity region in which points transition away from the analytically derived behaviour for cusp events for GWBs.

In order to adequately cover this region, we have chosen to randomly select points spanning the entire set of strings and binned them given that their respective velocities fell inside this region. Taking into consideration the previously arbitrarily defined lower bound for pseudocusps ($\log(1-v) = -3$), we chose to significantly increase the number of points spanning an order of magnitude either side of this limit ($-4 < \log(1-v) \leq -2$) in order to collect sufficient data for the yet undefined region of transition. Here we should note that by expanding our data in this way, we increase the number of points inside the velocity bounds for pseudocusps. However these points lie in the neighbourhood of cuspy events and must not be confused with independent pseudocusps. There is no such reason to expect these points to behave in a manor (see App.~\ref{sec:codeout}) different to any identified independent pseudocusp given the parameter of interest here is the velocity. Therefore these extended points offer no other purpose than to enrich the statistical accuracy of our analysis.

This gives a grand total of 10326 points we have analysed over the 119 strings which focus on the velocity range $-\infty < \log(1-v) \leq -2$ up to a numerical accuracy of $\mathcal{O}(10^{-6})$. Within these limits we can identify four velocity regions we wish to explore in order to sufficiently grasp the behaviour of the waveform for points on the string, namely cusps, pseudocusps, transitional and external. For the convenience of the reader, we repeat that we began with cusps {\it a priori} defined as points in the velocity range $-\infty < \log(1-v) \leq -6$ and pseudocusps as $-6 < \log(1-v) \leq -3$. We identify as transitional points any points falling in such a velocity region which leads to confined deviation from the expected analytical value, while any points yielding a non-relativistic behaviour will be labeled as external points. The velocity bound for such transitional and external regions have yet to be determined.

\subsection{Numerical simulation results}

\begin{figure}[t]
	\centering
	\includegraphics[width=1\linewidth]{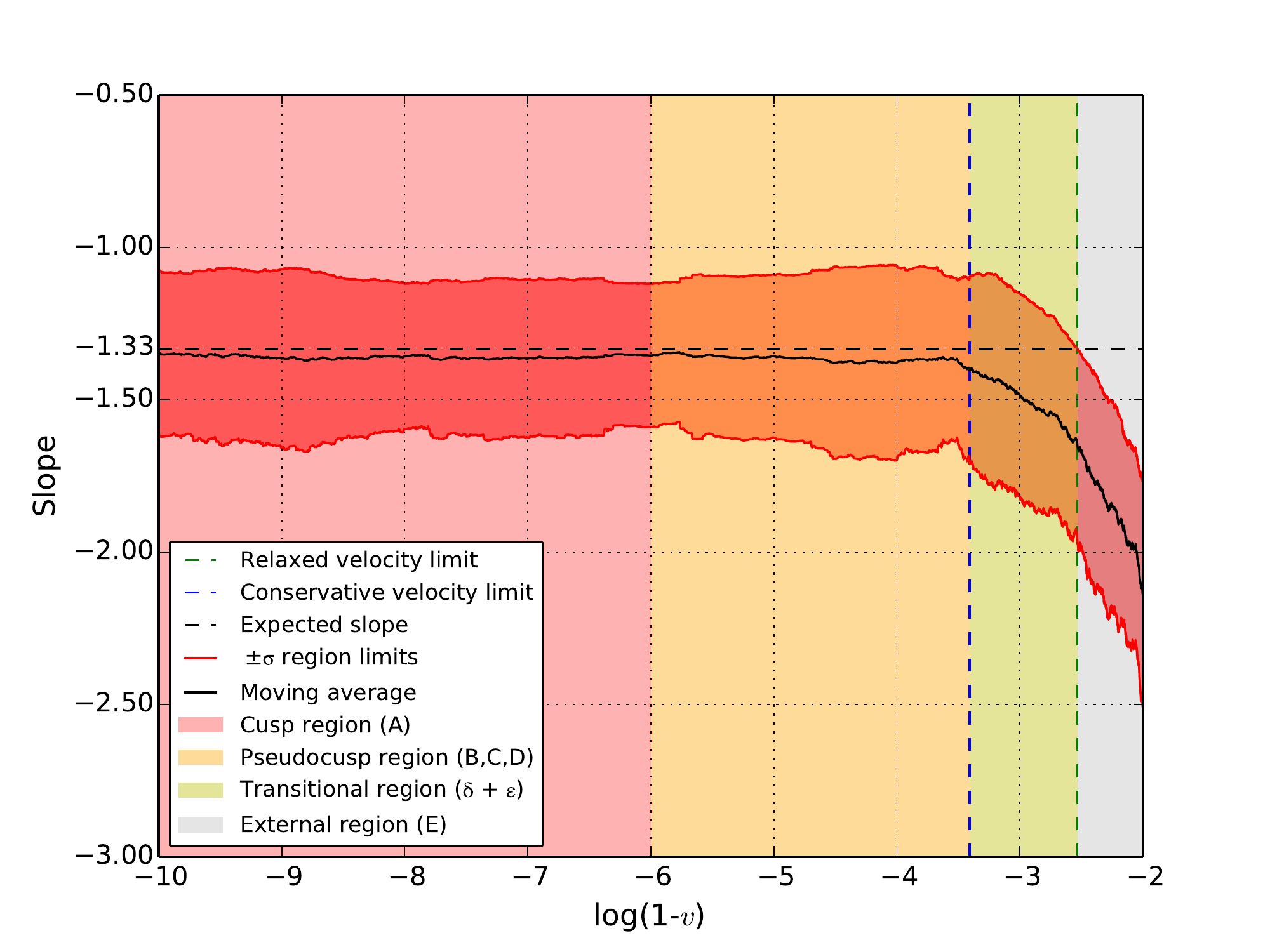}
	\caption{Calculated slope of the high frequency end of the GW waveform's power spectrum with respect to $\log(1-v)$ the logarithmic deviation of velocity from $c=1$, using the total 10326 points in the velocity range $1 > v \geq 0.99$. The convoluted moving point average is represented by the dot-dashed horizontal black line along with a $\pm \sigma$ deviation band in red. The velocity ranges are split into four regions, namely cusps (in red, for $\log(1-v) \in [-10,-6]$), pseudocusps (in orange, for $[-6,-3]$), external (in grey, for $[-3,-2]$) and transitional (in green, for $[-3.41,-2.53]$). The vertical dashed lines limit the transition region, giving a conservative and a relaxed threshold.}
	\label{fig:SlopeVsLogVel}
\end{figure}

In Fig.~\ref{fig:SlopeVsLogVel}, we present the total data for the calculated slope at the high frequency end of the GW waveform's power spectrum, from all 10326 points of the 119 strings, with respect to the logarithmic deviation of their velocity from $c=1$. We separated the velocity range into a cusp region (in red, for $\log(1-v) \in [-10,-6]$), a pseudocusp region (in orange, for $[-6,-3]$) and external region (in grey, for $[-3,-2]$). In addition to the arbitrary configured regions, subject to our analysis we also present a green region between the two vertical dashed lines shows, as we will detail below, a region for the threshold between highly relativistic points (of interest for the GWB emissions) and external points (surplus to requirements for this study). The convoluted moving average $\mu$ of the slope is represented by the solid black line, while the expected high velocity slope $\sfrac{-4}{3}$ is denoted by the horizontal black dot-dashed line. Finally, the red band gives the $1\sigma$ deviation from the average $\mu$.

First, one can see that in the high velocity limit, the convoluted moving average of the slope reaches the expected value of $ \sfrac{-4}{3}$, as expected from analytic computations. This is true for all points in the velocity range which defines them as cusps. Similarly, for low enough velocities (around $\log(1-v) \simeq -2$), the slope strongly deviates from such behaviour. We turn our attention to the region in between such limits as a focal point of interest in our results. Indeed, the expected high velocity behaviour remains exactly the same, down to almost all velocities defining pseudocusps (that is down to $\log(1-v) \simeq -3.5$ at least). This confirms that such points outside the formal definition of a cusp should be considered of interest in the same regard as actual cusps regarding their significance for GWB considerations.

As this high velocity behaviour carries on through the velocity region for pseudocusps, we would like to use this to define, accurately the limits on how these points are classified. Said differently, using the information in Fig.~\ref{fig:SlopeVsLogVel}, we can identify the limits in which we see a transition away from the GWB behaviour. First, one can remark that the {\it a priori} threshold, namely $\log(1-v) = -3$ lies roughly correct for a lower velocity bound. Still, we need to define an accurate way to draw this lower velocity bound, which we chose to extract under two different deliberations.

A conservative limit was set by seeking a ``significant'' deviation from the average behaviour at high velocities. Since the relative error of the calculated moving average, with respect to the expected, analytical behaviour $\sfrac{-4}{3}$, is roughly $\sim 2\%$ (on average in the velocity range $-10 < \log(1-v) \leq -4$, we defined $5\%$ as a significant deviation. \footnote{We only initially required that an average deviation should track for the limit for points defined as cusps ($-10 < \log(1-v) \leq -6$), although we find this track to velocities lower that the lower velocity bound for cusps ($-6 < \log(1-v) \leq -4$).} Such a constraint gives a conservative threshold of $v = 0.99961$, or $\log(1-v) = -3.41$, that is, an upper limit of the transition region. Our second approach is to consider a more relaxed limit coming from the crossing of $\mu+\sigma$ with the analytical value (where the dashed line exits the red band). This yields a lower transitional limit of $v = 0.99707$, or $\log(1-v) = -2.53$.

Using these two bounds we can now define a transitional region between $0.99961 > v \geq 0.99707$, or $-3.41 < \log(1-v) \leq -2.53$, where points could still exhibit some significant implications for GWB emissions but are ultimately beginning a transition away from the analytical behaviour of cusps. At this point, one can redefine the previous limits for pseudocusps to a velocity range of either $-6.0 < \log(1-v) \leq -3.41$ or $-6.0 < \log(1-v) \leq -2.53$. Alternatively, as the original bound lies roughly in the middle of the transition region, one could chose to keep it as $-6.0 < \log(1-v) \leq -3.0$ as a valid limitation. These considerations give a combined band of $- \infty < \log(1-v) \leq -3.41$ or $- \infty < \log(1-v) \leq -2.53$ for an inclusive region for cuspy events, referring to all points one would need to consider for GWB emissions.

\begin{figure}[t]
\begin{adjustwidth}{-5em}{-5em}
	\begin{center}
	\includegraphics[width=1\textwidth]{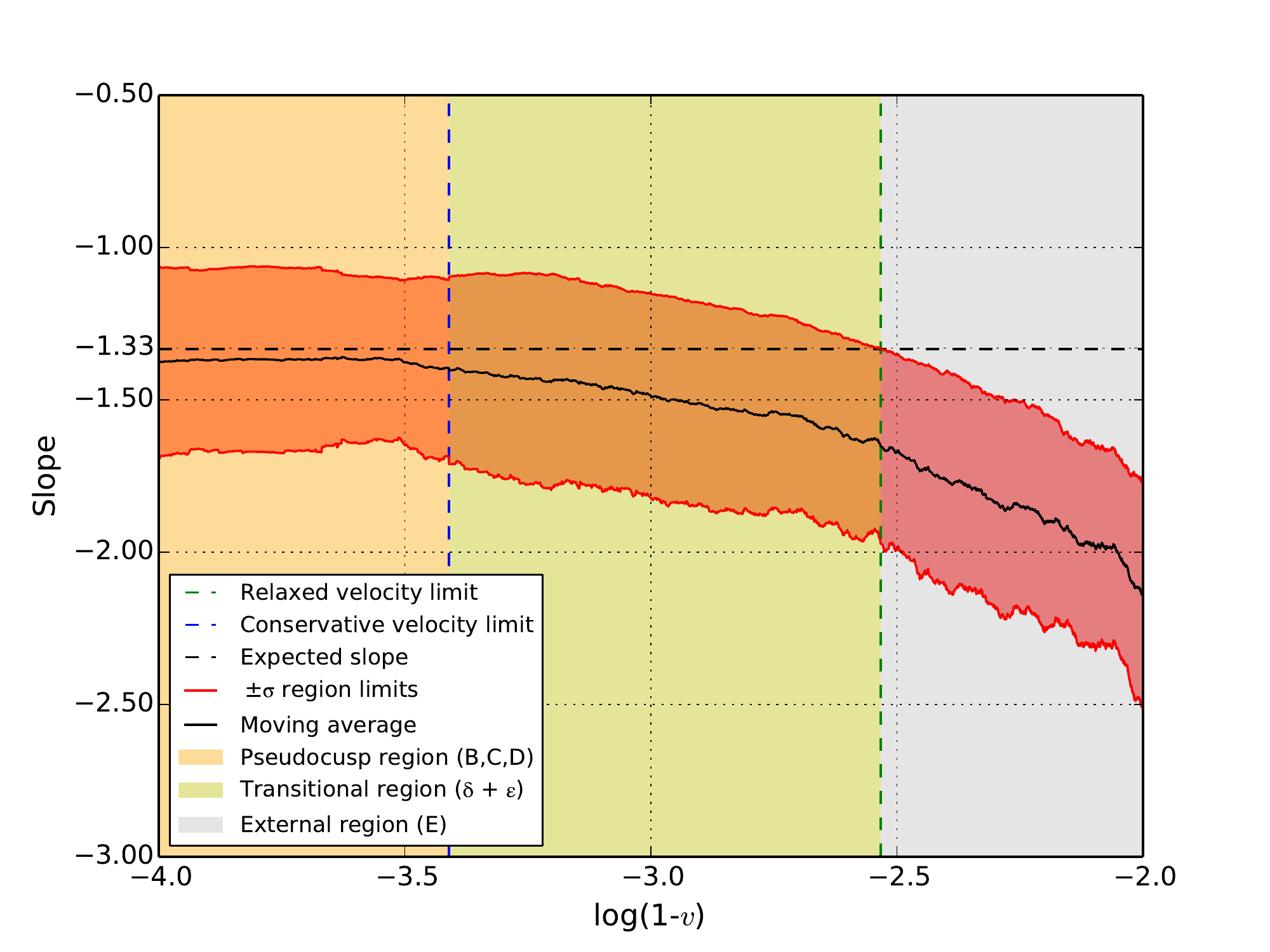}
	\end{center}
	\end{adjustwidth}
	\caption{Zoomed in region of Fig.~\ref{fig:SlopeVsLogVel}, where the high frequency behaviour transitions from its $\sfrac{-4}{3}$ analytical slope, away from the cusps behaviour.}
	\label{fig:SlopeVsLogVelZoom}
\end{figure}

For clarity, let us zoom into the region in which we identify the transition between the $\sfrac{-4}{3}$ frequency slope for the waveform to one decreasing as a function of velocity, as displayed in Fig.~\ref{fig:SlopeVsLogVelZoom}. One can clearly see how the various thresholds clearly define three regions: the high velocity end down to $\log(1-v) \lesssim -3.5$, the average and the $1\sigma$ band are horizontal and the GW emissions follow closely the analytical results for cusps; at the low velocity end up to $\log(1-v) \gtrsim -2.5$, the expected $\sfrac{-4}{3}$ result is far (and always further away) from the average, outside the $1\sigma$ band; in the medium range, roughly between $-3.5 \lesssim \log(1-v) \lesssim -2.5$, the average almost linearly deviates from the expected $\sfrac{-4}{3}$ value, but the $1\sigma$ band still contains it. This transitional region can be either excluded in a conservative way, included in a more relaxed approach, or approximately averaged via spliting in half, choosing $-3$ as the defining threshold. Note that this value lies approximately where the average deviates by about $10\%$ from the analytical value.

\subsection{Statistical distribution of string points}

\begin{figure}[t]
\begin{adjustwidth}{-3em}{-3em}
	\begin{center}
	\begin{subfigure}{9cm}
		\includegraphics[width=0.99\linewidth]{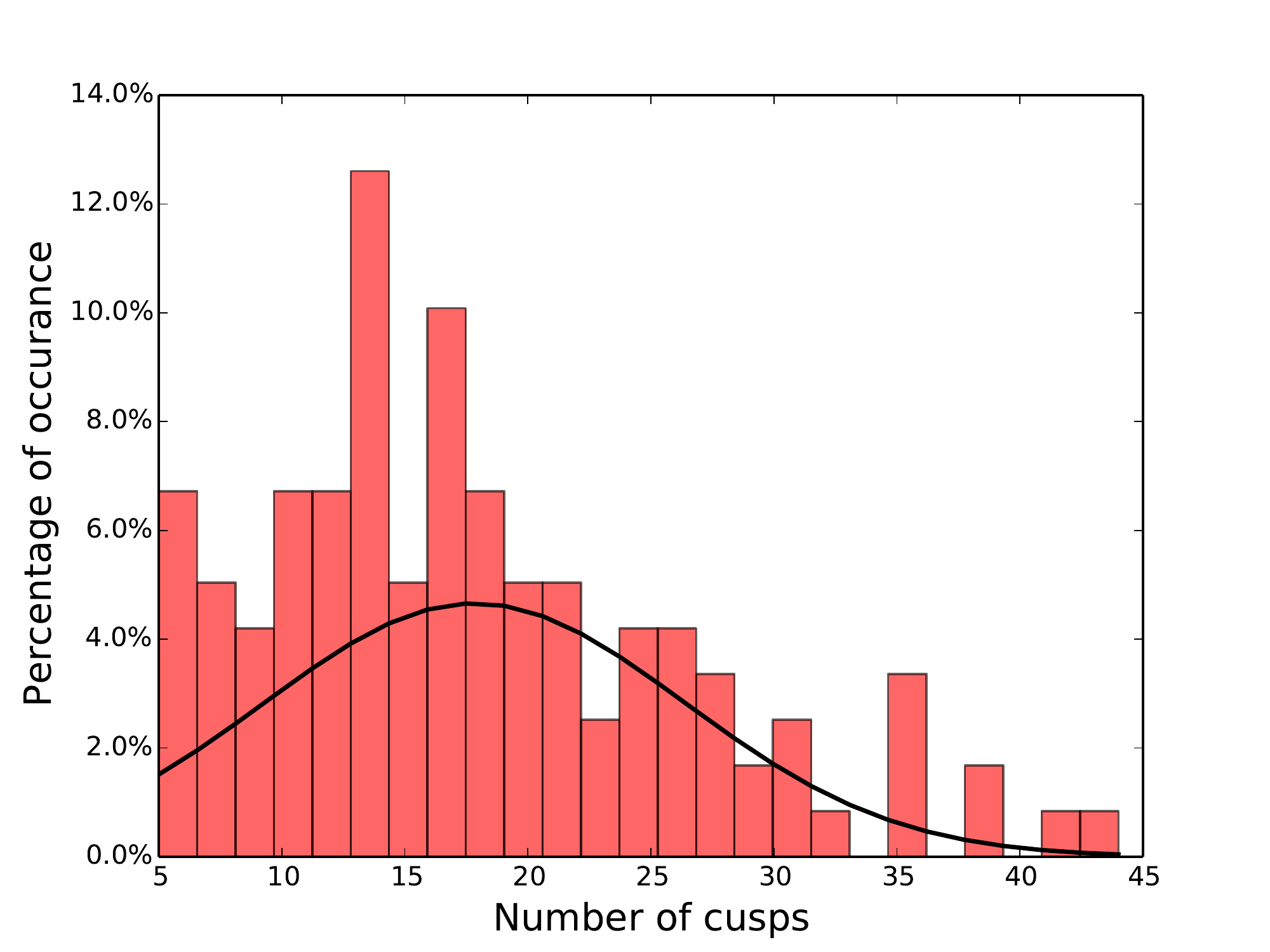}
		\caption{Number of cusps formed per string}
		\label{fig:nbCpString}
	\end{subfigure}
	\begin{subfigure}{9cm}
		\includegraphics[width=0.99\linewidth]{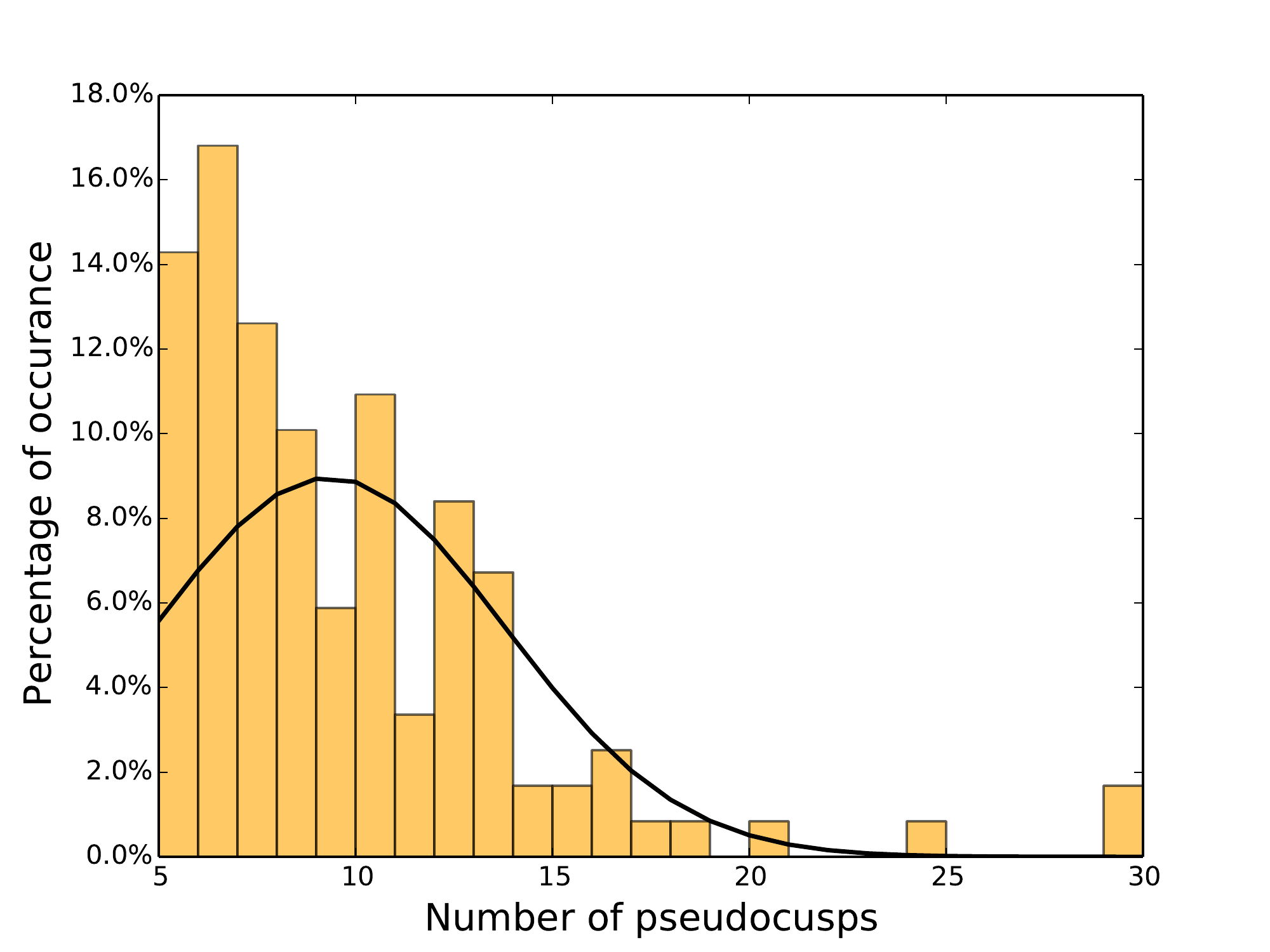}
		\caption{Number of pseudocusps formed per string}
		\label{fig:nbPCpString}
	\end{subfigure}
	\begin{subfigure}{9cm}
		\includegraphics[width=0.89\linewidth]{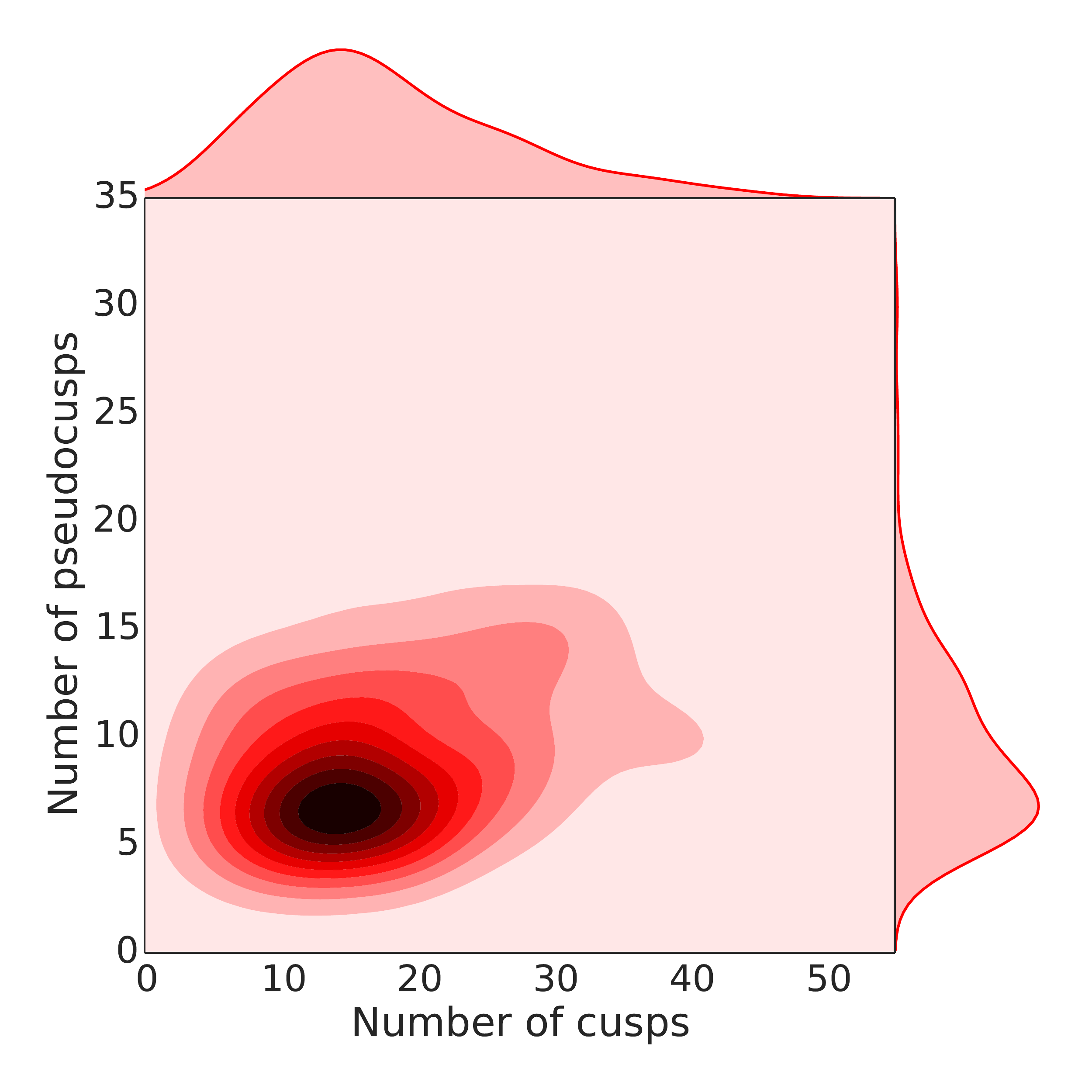}
		\caption{Contour density plot showing correlation between the number of cusps formed vs the number of pseudocusps formed per string along with corresponding density function plots.}
		\label{fig:contourdensity}
	\end{subfigure}
	\begin{subfigure}{9cm}
		\vspace{1em}
		\includegraphics[width=0.99\linewidth]{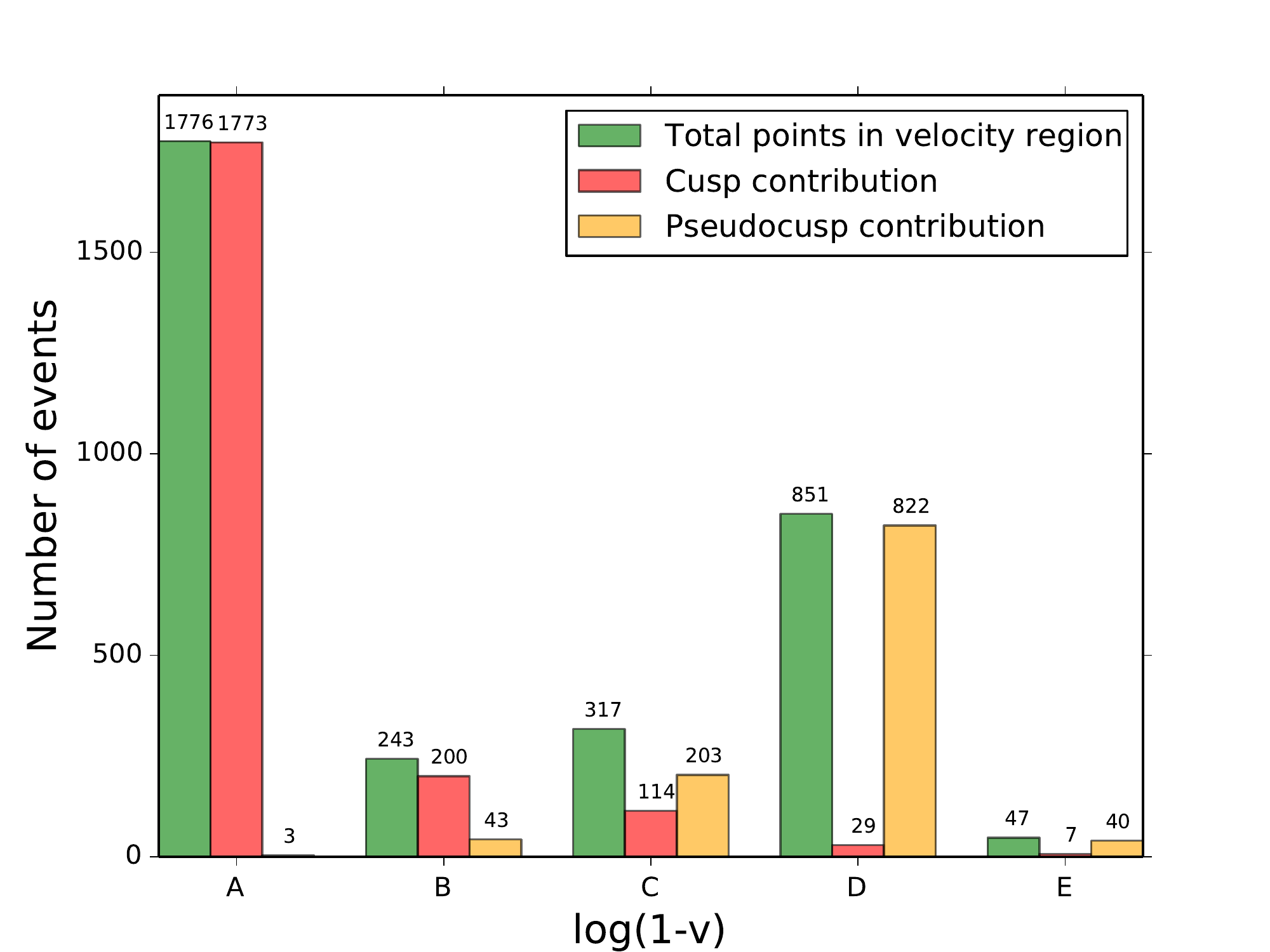}
		\caption{Bar charts representing the total number of events in each of the velocity regions defined in Table.~\ref{tab:percent} and their the corresponding contributions from both cusps and pseudocusps.}
		\label{fig:barchartCPC}
	\end{subfigure}
	\caption{Relationship plots for cusps and pseudocusps per string over the total of 119 strings\\
		highlighting the approximate 2:1 ratio relationship shared by cusps and pseudocusps respectfully\\
		as argued in \cite{ENS2014} (a),(b) and (c) along with the number of events contributing to each velocity\\
		region (d).}
	\label{fig:numcuspss}
	\end{center}
	\end{adjustwidth}
\end{figure}

To gain a good understanding of how such a change of the choice of velocity ranges included in a GWB computation could impact the potential output, Table~\ref{tab:percent} details the average percentage of the string moving in each velocity range of interest in our simulation. We divided the worldsheet of each string into $20000^2$ points (using a local inturpolted increase in the accuracy) in which we computed the velocity at each point and binned according to the following bounds. Region A corresponds to cusps, that is $- \infty < \log(1-v) \leq -6$, while ({\it a priori}) pseudocusps have been segregated into three regions, B, C and D, each one spanning one order of magnitude. Due to the fact we wanted to explore the possibility of changing the threshold with respect to the GW behaviour, we also considered a bin for velocities falling in $-3 < \log(1-v) \leq -2$, namely region E. Region F contains all the other points, which we expect to behave as non-relativistic points. The results found for each string have then been combined with the averages over the 119 strings presented.
\begin{table}[t]
	\begin{center}
	\begin{tabular}{cr@{$\;>v\geq\;$}lcc}
		\hline
		\multirow{2}{*}{Region}	& \multicolumn{2}{c}{\multirow{2}{*}{Velocity range}}	& Logarithmic		& Averaged percentage		\\
						& \multicolumn{2}{c}{}						& velocity range		& of the worldsheet	 (\%)		\\
		\hline
		A				& $1.0$		& $0.999999$				& $[- \infty, -6]$		& $0.000192$			\\
		B				& $0.999999$	& $0.999990$				& $[-6, -5]$			& $0.001651$				\\
		C				& $0.999990$	& $0.99990$					& $[-5, -4]$			& $0.016057$				\\
		D				& $0.99990$		& $0.9990$					& $[-4, -3]$			& $0.162562$				\\
		E				& $0.9990$		& $0.990$					& $[-3, -2]$			& $1.640088$				\\
		F				& $0.990$		& $0.0$					& $[-2, 0]$			& $98.179450$				\\
		\hline
	\end{tabular}
	\end{center}
	\caption{Averaged percentage of the worldsheet $\bar \Sigma \equiv \left\{ (\sigma_+,\sigma_-) \in [0, {\cal L}]^2 \right\}$ yielding a velocity within the intervals defining regions A to F. The percentages are computed on each string and then averaged over all 119 strings.}
	\label{tab:percent}
\end{table}
As we have stated, we chose to define the transitional region by both a conservative and relaxed approach, the former using the shift of the average from the expected analytical value and the latter when this expected value exits the standard deviation band. We then compared this to the previous arbitrarily set bound chosen for pseudocusps in Ref.~\cite{ENS2014}. In Table~\ref{tab:percent}, it is obvious that the velocity bound for a point approaching the limit $c=1$ can have a large impact on the selected proportions of the string to evaluate for GWB emissions. Interestingly, one can note that not only are cusps representing a very small fraction of the whole worldsheet (which is to be expected), but also of the highly relativistic patches. The important concern we wish to note is the relationship between the velocity constraints placed on points of interest and the fraction of the string we are considering. This may be taken into account in further studies of string network evolution. Indeed, the proportion of the string moving at velocities above $0.999$ (namely points within regions A to D) produces an order $\mathcal{O}(10^3)$ enhancement to the proportion of the string we could otherwise consider. With independent pseudocusps appearing at an order of $\mathcal{O}(1)-\mathcal{O}(10)$ per string as shown in Fig.~\ref{fig:nbPCpString} and the total number of points above the pseudocusp lower velocity bounds (see Table \ref{tab:percenttwo}) representing a far greater number of points on the string, this indicates that independent pseudocusps will only represent a rather small fraction of the total number of pseudocusps. More importantly, it also implies that considering only cusps and independent pseudocusps provides an extremely limiting approach with respect to considering the emission of high frequency GWs from cosmic strings.

In addition, one can look at how the modification of the bound (using the {\it a priori} fixed limit $\log(1-v) = -3$, the conservative limit $\log(1-v) = -3.41$ or the relaxed limit $\log(1-v) = -2.53$) would impact such proportions. Table~\ref{tab:percenttwo} details the averaged percentage of the worldsheet in newly defined regions taking into account our findings from Fig.~\ref{fig:SlopeVsLogVel}. Regions $\delta$ and $\varepsilon$ detail the corresponding variations of the considered proportion of worldsheet that different velocity cutoffs play with respect to the previously defined arbitrary limits. The former encompasses velocities between the conservative bound ($\log(1-v) = -3.41$) and the {\it a priori} bound ($\log(1-v) = -3$), while the latter contains velocities between the {\it a priori} bound and the relaxed bound ($\log(1-v) = -2.53$). Alternatively, region $\delta$ details the intersection of the {\it a priori} defined pseudocusp regions (B, C and D) and the transition region, while region $\varepsilon$ is the intersection between the external region (E) and the transition region. Regions ${\bar\delta}$ and ${\bar\varepsilon}$ are respectively the complement of $\delta$ and $\varepsilon$ within regions D and E (meaning that ${\bar\delta} \cup \delta = D$ and ${\bar\varepsilon} \cup \varepsilon = E$).
\begin{table}[t]
	\begin{center}
	\begin{tabular}{rlr@{$\;>v\geq\;$}lr@{$,\,$}lc}
		\hline
		\multicolumn{2}{c}{\multirow{2}{*}{Region}}				& \multicolumn{2}{c}{\multirow{2}{*}{Velocity range}}	& \multicolumn{2}{c}{Logarithmic}	& Averaged percentage	\\
		\multicolumn{2}{c}{}							& \multicolumn{2}{c}{}						& \multicolumn{2}{c}{velocity range}	& of the worldsheet	 (\%)	\\
		\hline
		\multirow{2}{*}{D \;\Bigg\{}	& ${\bar\delta}$			& $0.99990$		& $0.99961$					& $[-4.0$		& $-3.41]$		& $0.052367$			\\
							& $\delta$				& $0.99961$		& $0.9990$					& $[-3.41$		& $-3.0]$		& $0.110196$			\\
		\multirow{2}{*}{E \;\Bigg\{}	& $\varepsilon$			& $0.9990$		& $0.99707$					& $[-3.0$		& $-2.53]$		& $0.348867$			\\
							& ${\bar\varepsilon}$		& $0.99707$		& $0.990$					& $[-2.53$		& $-2.0]$		& $1.291221$			\\
		\arrayrulecolor{gray}\hline
		\multicolumn{2}{c}{$A \cup B \cup C \cup {\bar\delta}$}		& $1.0$		& $0.99961$					& $[- \infty$		& $-3.41]$		& $0.070267$			\\
		\multicolumn{2}{c}{$A \cup B \cup C \cup D$}				& $1.0$		& $0.9990$					& $[- \infty$		& $-3.0]$		& $0.180462$			\\
		\multicolumn{2}{c}{$A \cup B \cup C \cup D \cup \varepsilon$}	& $1.0$		& $0.99707$					& $[- \infty$		& $-2.53]$		& $0.529329$			\\
		\arrayrulecolor{black}\hline
	\end{tabular}
	\end{center}
	\caption{Averaged percentage of the worldsheet $\bar \Sigma \equiv \left\{ (\sigma_+,\sigma_-) \in [0, {\cal L}]^2 \right\}$ yielding a velocity within new intervals, considering the results from Fig.~\ref{fig:SlopeVsLogVel}. We define regions $\delta$ and $\varepsilon$ as the transition regions, and their (respective) complement $\bar \delta$ and $\bar \varepsilon$ relative to (respectively) $D$ and $E$. For each choice of the threshold, namely $\log(1-v) \in \{-3.41, -3.0, -2.53\}$, we also give the physically relevant percentages of cuspy events. The percentages are computed on each string and then averaged over all 119 strings.}
	\label{tab:percenttwo}
\end{table}

These regions, along with their complements, can be used to define three areas which encode the total number of points leading to important GWB contributions, hence providing a behaviour indicative of cuspy events on cosmic strings. Indeed, while choosing the {\it a priori} bound would impose, for GWB analyses, selecting all points within region $A \cup B \cup C \cup D$, choosing the conservative (relaxed) bound implies selecting points within $A \cup B \cup C \cup {\bar\delta}$ (respectively $A \cup B \cup C \cup D \cup \varepsilon$). Due to the bend the transitional region exhibits away from cusp behaviour, we find this grants an allowance for an $\mathcal{O}(10^2)$ enhancement to the fraction of the string we could consider of interest with either relaxed or conservative bounds. It is worth noting that in regions $\delta$ and $\varepsilon$, we find a $-0.110\%$ or $+0.349\%$ increase to the considered worldsheet with respect to the {\it a priori}, arbitrary limit $\log(1-v) = -3$. Alternatively, this means a factor $2$ to $3$ enhancement between the worldsheet proportions defined by each bound. We believe this to be insignificant compared to the factor $10^2$ to $10^3$ gained with respect to the cusp-only bound.

Let us emphasise that following analytical studies of Ref.~\cite{Turok:1984cn} for example, one may conclude that there is $\mathcal{O}(1)$ cusp per string. The purpose of our study was not to quantify the number of occurring cusps, since this would depend on the parameters of a {\sl natural} string network, but to give an order of magnitude estimate of the number of points (per cuspy event) on the string's worldsheet that can play the same role as cusps with respect to the emission of GWBs. In Ref.~\cite{ENS2014}, the number of cusps in the numerical setup was quite high (see Fig.~\ref{fig:nbCpString}) with the number of near crossings, or independent pseudocusps representing approximately half the number of cusps (see Fig.~\ref{fig:nbPCpString}). Figure~\ref{fig:contourdensity} shows the contour density plot for the number of cusps and independent pseudocusps we identified, giving a relatively good realisation of this relationship in our simulation. Of course, this approximate trend depends highly on the velocity thresholds for a point to be labeled as either a pseudocusp or as a cusp\footnote{Recall in Ref.~\cite{ENS2014}, due to numerical inaccuracies, points were labeled as cusps as soon as $-\infty < \log(1-v) \leq -6$.} but as we have seen, the cuspy event bound $\log(1-v) < -3$ is {\it a posteriori} suitable. As we are predominantly interested in the properties of pseudocusps, we wanted to ensure we have analysed a significant number of points with a velocity presenting a significant distance away from the limits between regions in order to counteract possible numerical uncertainties in our code. That is, we demand a sizeable fraction of the number of events to fill each of the regions A to D in Table~\ref{tab:percent}, as can be seen from Fig.~\ref{fig:barchartCPC}.

A final note on our method and numerical inaccuracies. For any direct crossing in which we calculate the velocity of the point to be superluminal, such error comes from the initial numerical accuracy of the string position vector. As previously stated, any point, including a cusp, will satisfy $|{\bf \dot{X}}_+| = 1 = |{\bf \dot{X}}_-|$. The numerical uncertainty in our code however can allow for fractional variations away from a unitary value which in turn allows for superluminal velocities below the numerical accuracy of $1.0 + 10^{-6}$. To correct these very minor offsets we allow for a normalisation of the velocity with respect to its norm defined as
\be
	{\bf \dot{X}} \rightarrow \frac{\bf \dot{X}}{|\bf{\dot{X}}_{\pm}|} ~.
\ee
It is worth noting that this logic also extends to the lower limit, such that direct crossings can have velocities just inside the upper pseudocusp limit.\footnote{This means we will have a small difference in the number of points we have identified as direct crossings (formal cusp) and the number of points present in the cusp velocity band. This plays no significant role in our conclusions given these points do not lie in a transitional region anyway. These discrepancies are shown in Fig.~\ref{fig:barchartCPC} which details the number of formally defined points which contribute to each velocity region defined in Table \ref{tab:percent}.}

The important aspect of our study and the global picture presented in Fig.~\ref{fig:SlopeVsLogVel} is that any points of importance for GWBs should be classified as a function of their velocity and not just the nature of the encounter between the ${\bf \dot X}_+$ and $-{\bf \dot X}_-$ curves. Certainly, it is also clear that so called independent pseudocusps from near crossings produce the same behaviour as the analytical predictions for cusps, with some freedom as to how far this behaviour tracks with respect to their velocity parameter. Said differently, both independent pseudocusps and those in the neighbourhood of a cusp must be accounted for by allowing fractional deviations of the velocities of any point of interest away from the luminal limit $c=1$. Such results from our study are of importance as this would lead us to consider a significant increase in the number of points when computing the GWB output given they adhere to certain velocity constraints. This enhancement to the proportion of the worldsheet we should look at, when incorporating the lowest acceptable velocities, that is, for points in regions $A \cup B \cup C \cup D \cup \varepsilon$ as apposed to just direct crossings, can account to a factor enhancment of $\mathcal{O}(10^3)$.

\section{Conclusions} \label{sec:conclusions}

The overarching message from our study has been that the consideration of points classified as pseudocusps can provide a significant enhancement (namely $\mathcal{O}(10^3)$) to the fractional percentage of the worldsheet one might choose to examine as important for GWB emissions from cosmic strings. The consideration of independent pseudocusps provides approximately a 50\% enhancement to the number of points to consider over just looking at cusps. Certainly then we have shown that there is a large increase in the number of points to consider when we extend these considerations to include pseudocusps in the vicinity of cuspy events. We have considered the non-interacting movement coming from light strings stretched between two junctions with fixed heavy strings, where by we assume our qualitative results remain valid more realistic string configurations. Analysis of the set of 119 strings allowed us to identify 2123 points classified as cusps, that is, points reaching the speed of light $c=1$ in relation to a crossing of the ${\bf \dot X}_+$ and $-{\bf \dot X}_-$ curves on the unit sphere, and approximately half as many, 1111 to be precise, classified as independent pseudocusps, that is, points reaching highly relativistic velocities but not exactly $c=1$, due to the close approach of the ${\bf \dot X}_+$ and $-{\bf \dot X}_-$ curves.

It was first shown analytically in Ref.~\cite{Damour2000} that (exact) cusps on cosmic strings present a decreasing power law at the high frequency end of the power spectrum which amounts to $f^{-4/3}$. They only incorporated the treatment of exact cusps but not two classes of highly relativistic points, corresponding to either an independent close approach of the ${\bf \dot X}_+$ and $-{\bf \dot X}_-$ curves or points in the neighbourhood of cusp which are slightly displaced either spatially or temporally. We have firstly clarified that the nature of such highly relativistic points, designated as pseudocusps, does in fact follow the same analytical behaviour and imply the same physical significance presenting a power law dependence equivalent to cusps. These points were initially designated by velocity limits $-6 < \log(1-v) \leq -3$ and as such we found that any point within this range exhibited a behaviour much the same as cusps.

The second phase of our study was to determine the fractional magnitude from the limit $c=1$ to which points begin to deviate from points classified as important for GWB emission considerations. To do this we analysed the limits coming from three regions: cusps, pseudocusps and transitional points along with points external to this lower limit, for consistency. In doing so we were able to both reinforce the importance of highly relativistic points, along with points which may maintain some of this importance (transitional), but also reclassify the velocity thresholds in which such points appear within our numerical environment. In particular we showed that points to be considered for GWBs should be determined by their velocity, relaxing the concept of just focusing on points moving at exactly the speed of light.

From the results of our study we arrive at a classification for a set of points encompassing an importance regarding high frequency GWBs whose velocity lies roughly within $-\infty < \log(1-v) \leq -3$, accounting for about $0.18\%$ of the worldsheet in our case (to be compared with the $0.00019\%$ if one considers only $-\infty < \log(1-v) \leq -6$, that is, a factor ${\cal O}(10^3)$). One could select a more conservative velocity limit $-\infty < \log(1-v) \leq -3.41$, implying reducing the size of the considered part of worldsheet by a factor $2$ to $3$. Alternatively, one could choose a more relaxed one, namely $-\infty < \log(1-v) \leq -2.53$, leading to a gain of such part's size by a factor $2$ to $3$.

In addition to the possible higher number of cusps per string (than the analytical ${\cal O}(1)$), this means that the high frequency gravitational waves output from a string network can be significantly enhanced with respect to previous computations. An interesting follow up would be to translate this work to incorporate the bounds on the strings' and network's parameters, in particular the tension $G \mu$.

\appendix
\section{Visualisation of points of importance for GWBs}

We seek here to represent, to provide a clearer understanding to the reader, the cuspy events and their environment, both in the unit sphere description as well as on the string.

\begin{figure}[t]
\begin{adjustwidth}{-3em}{-3em}
	\begin{center}
	\begin{subfigure}{9cm}
		\includegraphics[width=0.8\linewidth, trim=210 250 120 140, clip]{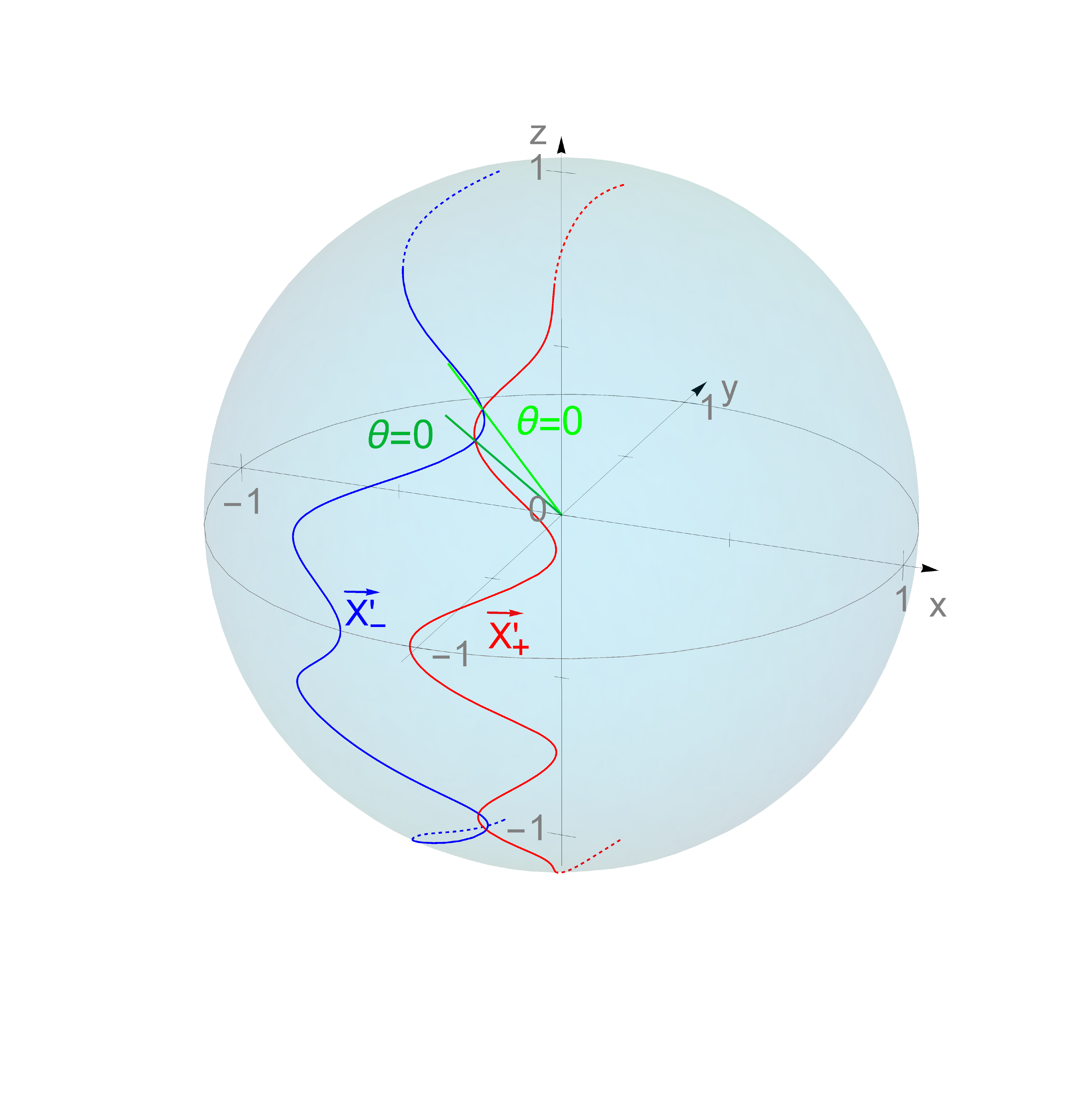}
		\caption{Representation of a (pair of) crossing(s) of the two curves identifying the formation of a (pair of) cusp(s). The angle $\theta$ is null at the crossing and becomes (algebraically) negative between them.}
		\label{fig:UnitSphereCusps}
	\end{subfigure}
	\begin{subfigure}{9cm}
		\includegraphics[width=0.8\linewidth, trim=210 250 120 140, clip]{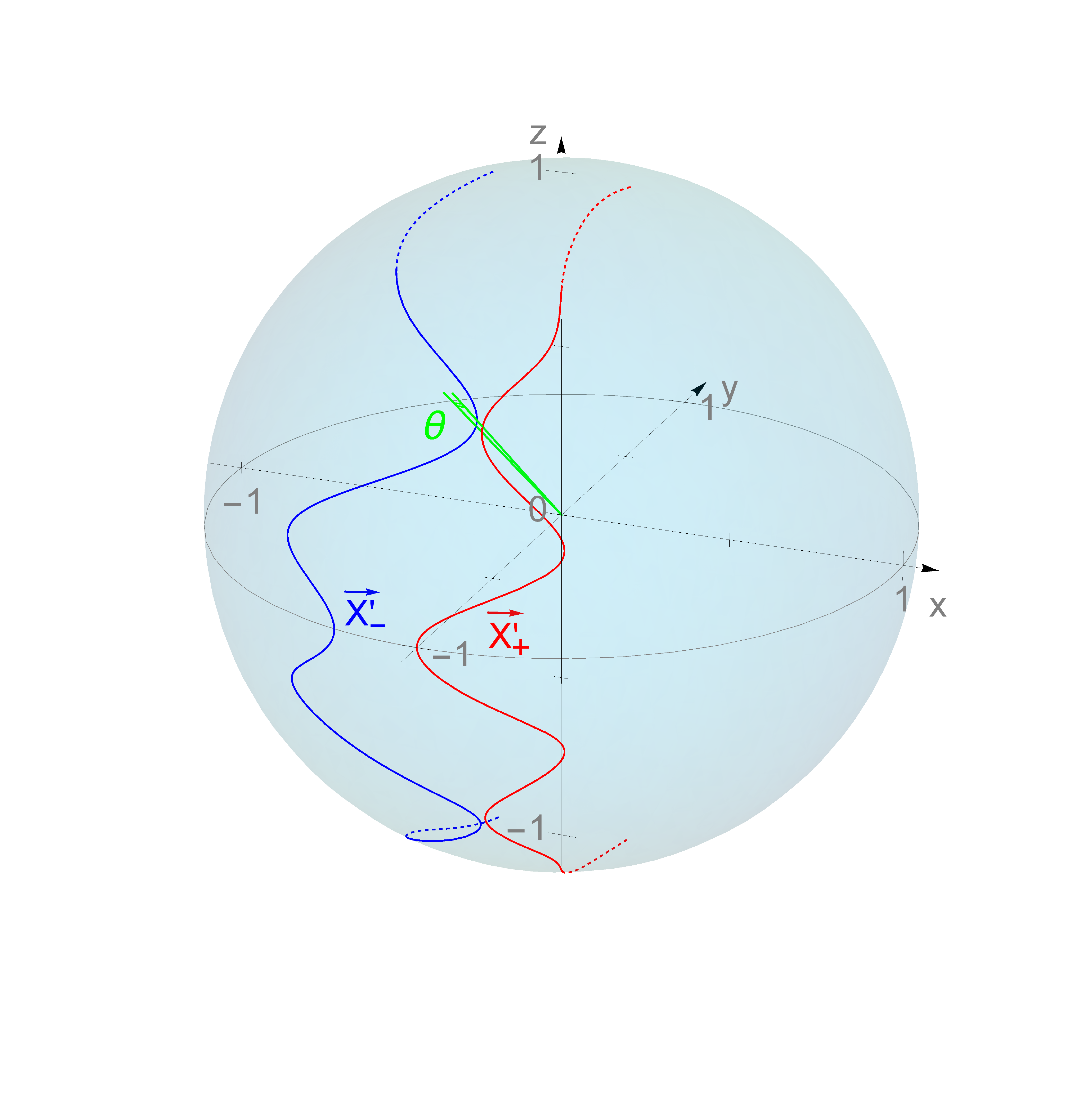}
		\caption{Representation of a near crossing of the two curves identifying the formation of an independent pseudocusp. The angle $\theta$ measures the softness of the pseudocusp, that is, the deviation from an exact (pair of) cusp(s).}
		\label{fig:UnitSpherePC}
	\end{subfigure}
	\end{center}
	\end{adjustwidth}
	\caption{Representations of close approach of the ${\bf \dot X}_+$ and ${\bf \dot X}_-$ curves on the unit sphere, used to identify cusps (direct crossings) and pseudocusps (near crossings) respectfully.}
	\label{fig:UnitSphere}
\end{figure}

Figure~\ref{fig:UnitSphere} shows representations of events on the unit sphere of the ${\bf \dot X}_+$ and ${\bf \dot X}_-$ curves in the case of cusps and pseudocusps. While one can see the crossings of the curves in Fig.~\ref{fig:UnitSphereCusps} and note the separation angle $\theta = 0$ at the cusps, Fig.~\ref{fig:UnitSpherePC} shows their close approach without any crossing, $\theta$ remaining non-null (and positive). Recall that it is the equality between such vectors which allows, in the GW emission computations, to cancel the leading term by choosing the emission direction aligned with such vectors and thus to enhance the high frequency end of the power spectrum, therefore leading to a GWB. Note that in the neighbourhood of a cusp, the vectors ${\bf \dot X}_+$ and ${\bf \dot X}$ are very close but not equal, implying a region of the string, in the vicinity of the cusp, which travels at highly relativistic velocities.

\begin{figure}[t]
	\begin{center}
	\includegraphics[width=0.7\textwidth, trim=120 50 40 40, clip]{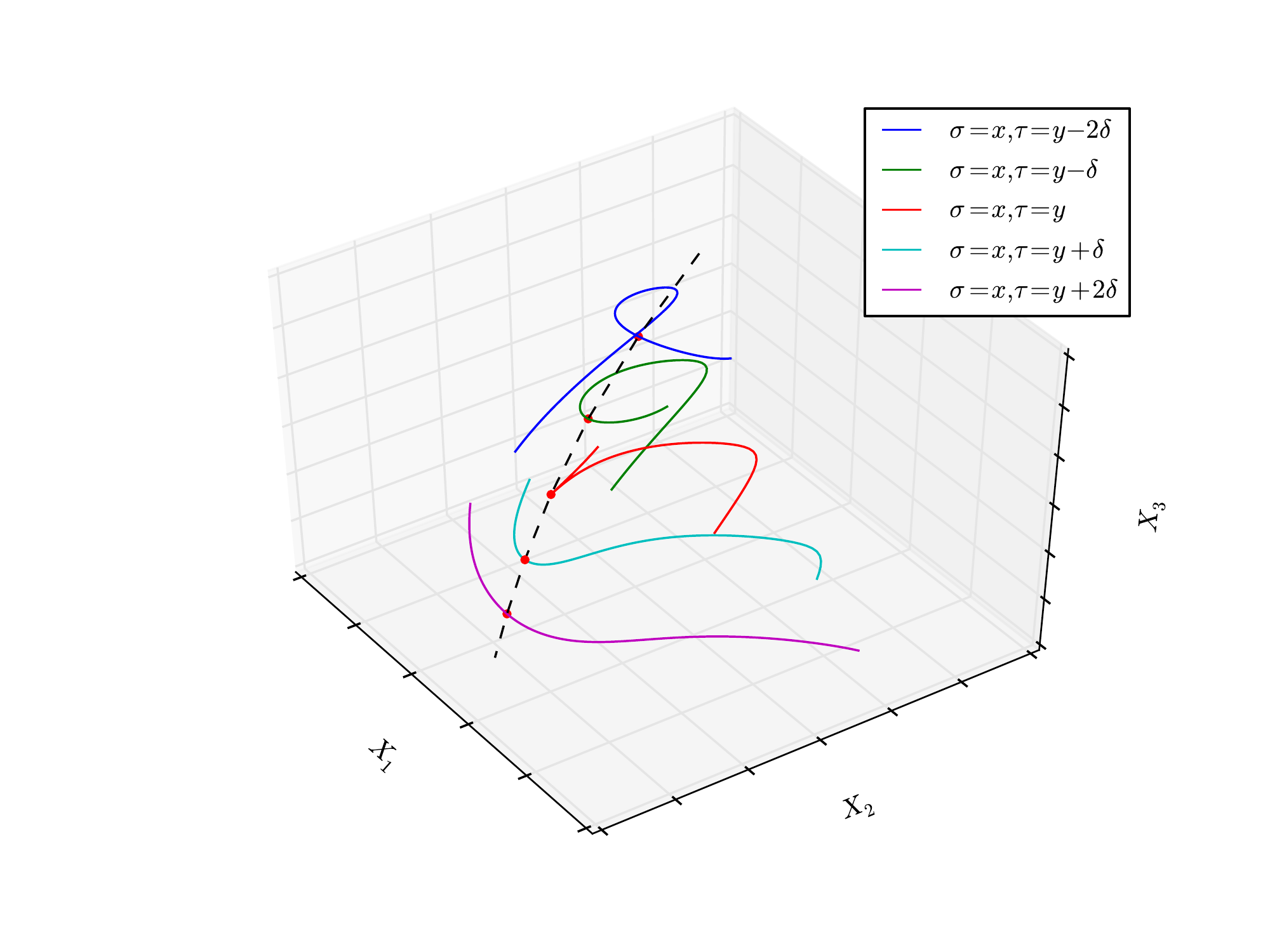}
	\caption{Simulation of a cusp's formation, with snapshots of a section of the string at different instants both before and after the formation of the cusp. The dotted line represents the time evolution of the point $\sigma = x$ on the string which becomes a cusp at the time $\tau = y$.}
	\label{fig:SnapshotsCusp}
	\end{center}
\end{figure}

As we mentioned already, a cusp, characterised by ${\bf \dot X} = 1$ and ${\bf X'} = 0$ (thanks to the Virasoro condition Eq.~(\ref{eq:Vira1})), is a point of infinite acceleration and null radius of curvature. This is shown in Fig.~\ref{fig:SnapshotsCusp} where the immediate vicinity $\sigma \in [x-\bar\delta, x+\bar\delta]$ of a cusp on the string is simulated and given at different instants, both before and after the formation of the cusp itself. The dotted line allows to track the movement of the point $\sigma=x$ which becomes the cusp at the instant $\tau = y$. One can clearly see the spiky shape of the string at the cusp as well as the continuous deformation which leads to such event.

\section{Numerical simulation output examples} \label{sec:codeout}

In this section we provide a few outputs from our numerical simulation at various stages, providing details on a few key stages of the processes involved in our study.

\begin{figure}[t]
	\begin{adjustwidth}{-6em}{-3em} \begin{center}
	\begin{subfigure}{17cm}
	\includegraphics[width=0.97\textwidth]{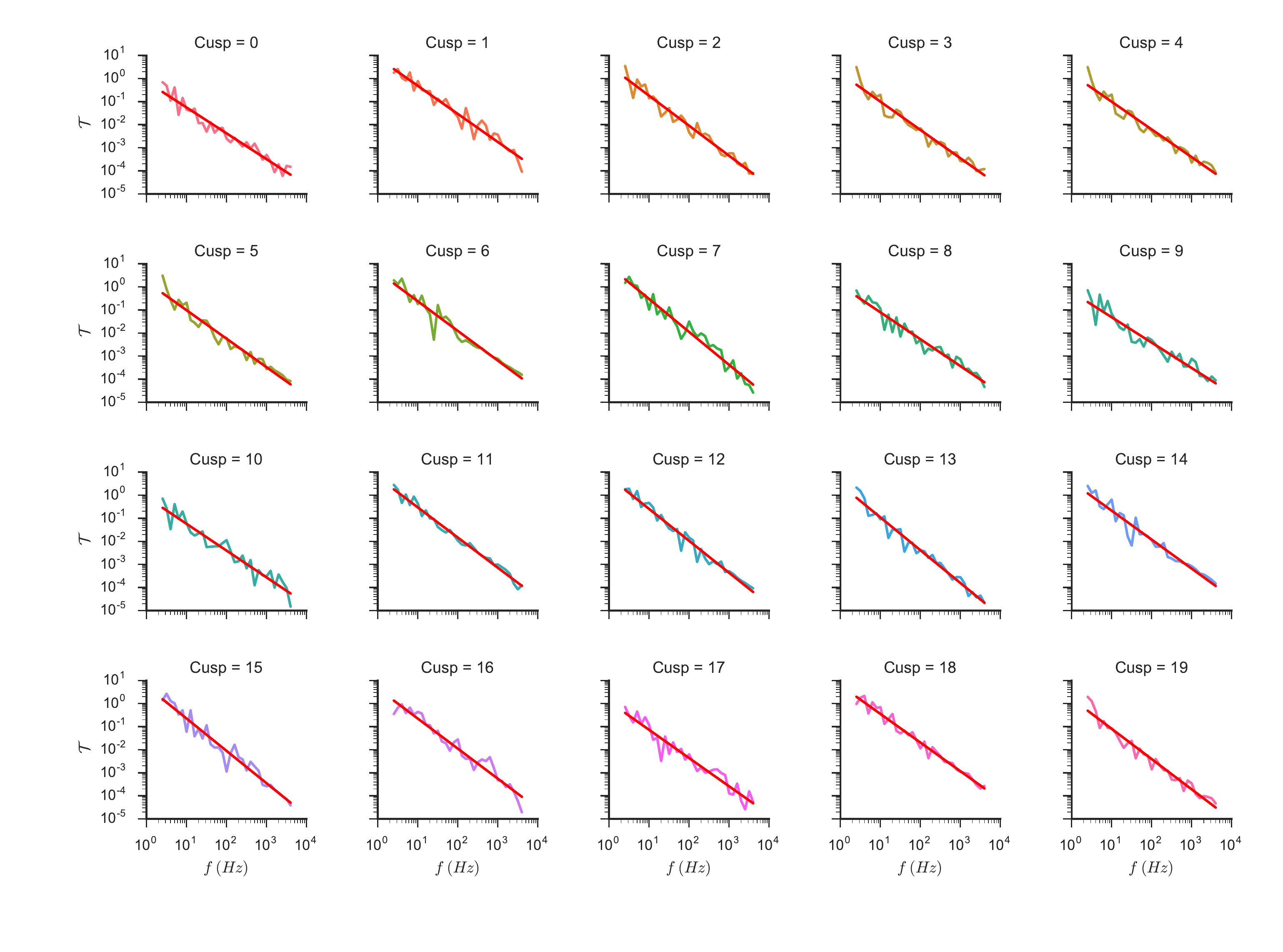}
	\end{subfigure}
	\end{center}
	\vspace{-2em}
\end{adjustwidth}
	\caption{High frequency window of the GW power spectrum for all the cusps (20) of a random string, as well as the linear regression yielding the calculated slope, in red. Only 26 frequency points in the frequency range $[2.52, 1000]~{\rm Hz}$ have been used.}
	\label{fig:string1plots}
	\vspace{1.5em}
	\begin{adjustwidth}{-4em}{-4em}
	\begin{center}
	\begin{subfigure}{9cm}
		\centering
		\includegraphics[width=0.99\linewidth]{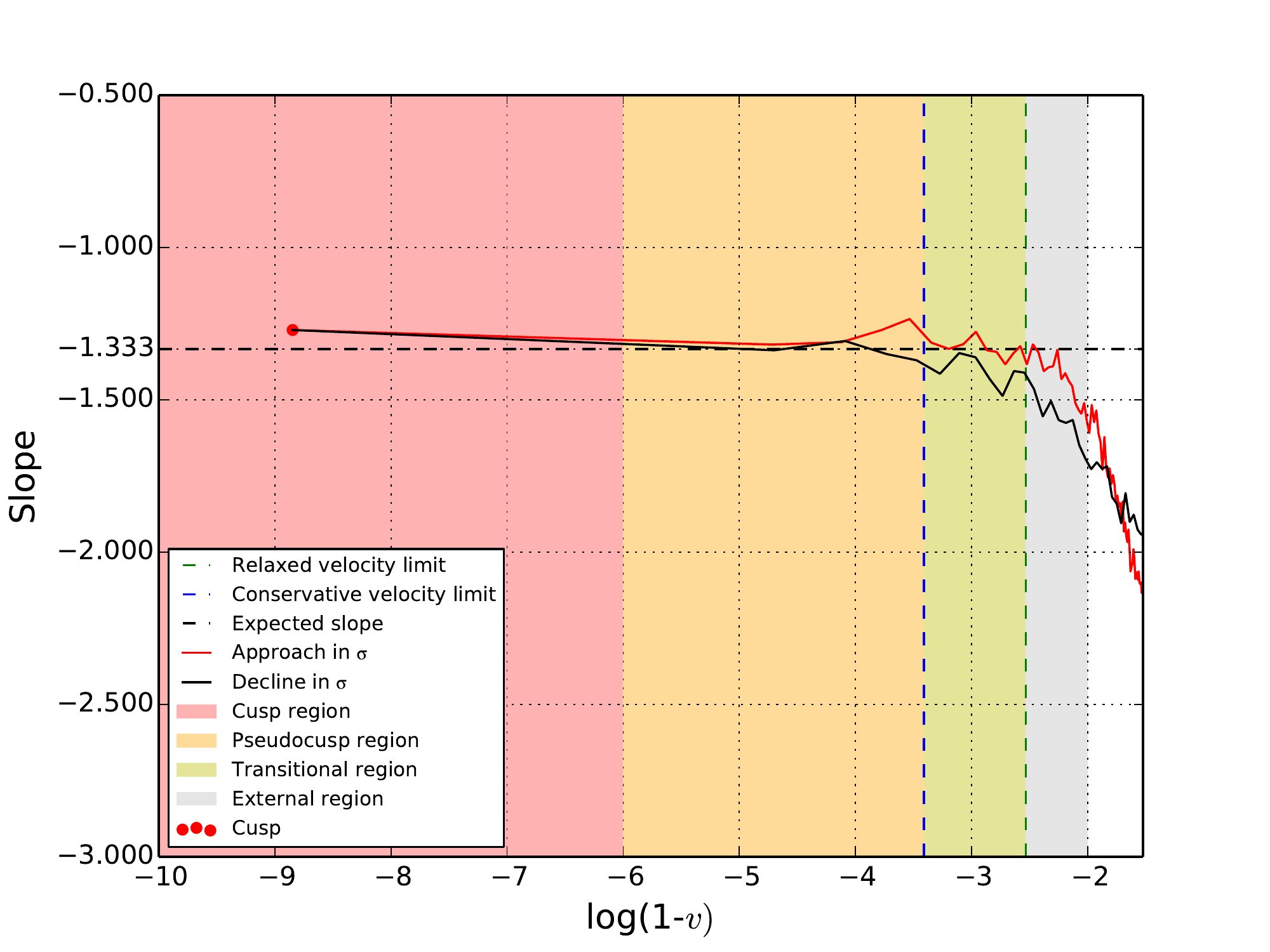}
		\caption{Spacial evolution in $\sigma$}
		\label{fig:SlopeCuspVicinitySig}
	\end{subfigure}
	\begin{subfigure}{9cm}
		\centering
		\includegraphics[width=0.99\linewidth]{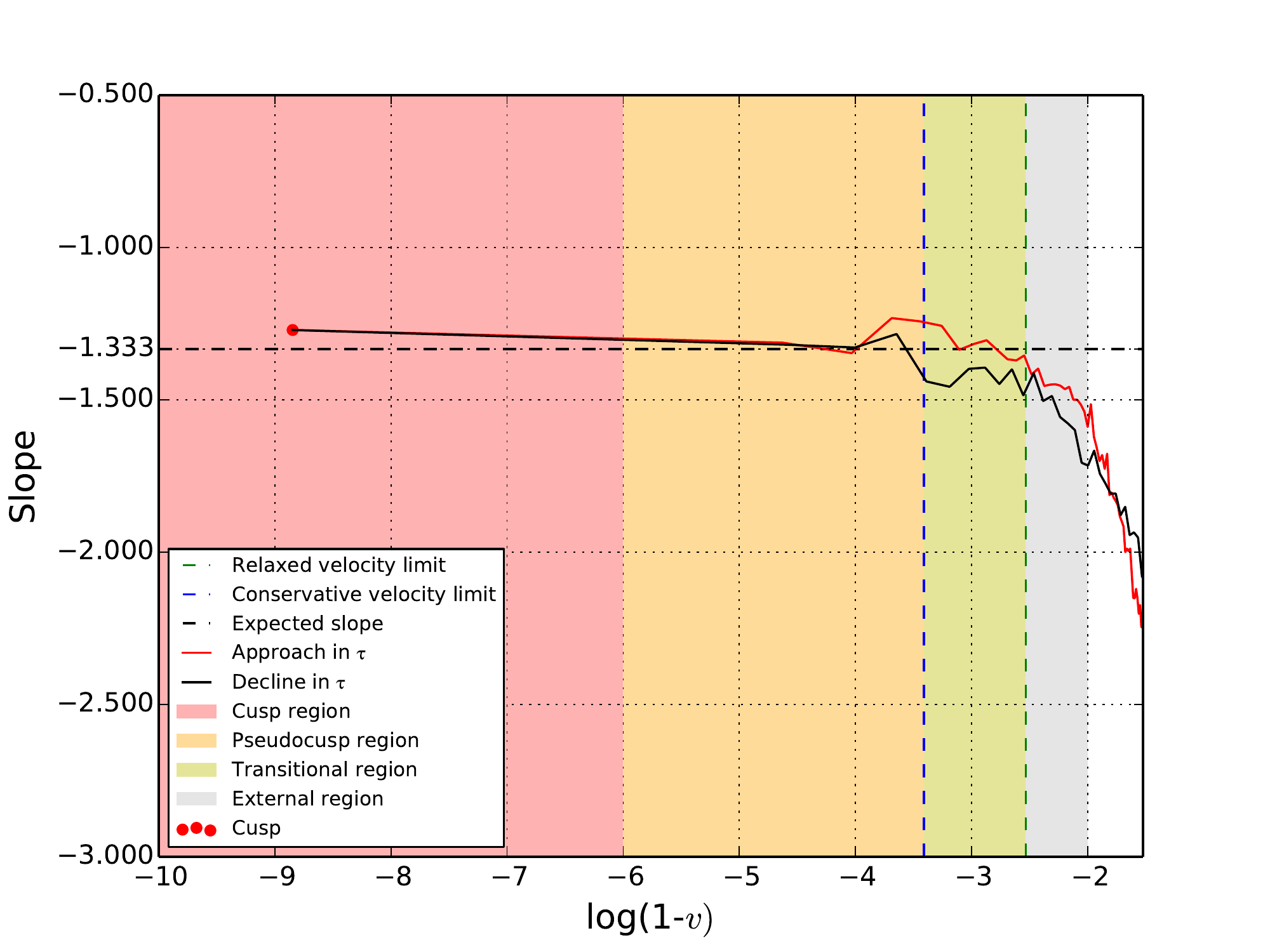}
		\caption{Temporal evolution in $\tau$}
		\label{fig:SlopeCuspVicinityTau}
	\end{subfigure}
	\end{center}
	\end{adjustwidth}
	\caption{Calculated slope of the GW power spectrum with respect to $\log(1-v)$ the logarithmic velocity, in the neighbourhood of a cusp. The velocity ranges, vertical and horizontal dashed lines are defined as in Fig.~\ref{fig:SlopeVsLogVel}. The red dot represents the cusp event, while the red (black) curve shows the approach to (respectively, the decline out of) the cusp, that is $\sigma < \sigma^{\rm (c)}$ or $\tau < t^{\rm (c)}$ (resp. $\sigma > \sigma^{\rm (c)}$ or $\tau > t^{\rm (c)}$). (a) yields the behaviour of the vicinity of the cusp in the spatial $\sigma$ direction; (b) yields the same in the temporal $\tau$ direction.}
	\label{fig:SlopeCuspVicinity}
\end{figure}

Figure~\ref{fig:string1plots} provides, for each cusp of one of our strings, the (26 points) high frequency window of the GW power spectrum as well as the calculated slope's linear regression in red. As one can see in a glance, our choice of frequency range provided consistent behaviour for such a panel of cusps. This is the treatment we imposed on each of the cusps found on each string of our simulation, leading to Figs.~\ref{fig:SlopeVsLogVel} and~\ref{fig:SlopeVsLogVelZoom}.

In addition, Fig.~\ref{fig:SlopeCuspVicinity} presents, for a random cusp, the slope obtained when considering its immediate environment. Indeed, while the cusp itself is represented by the far left point in red (with velocity around $1-10^{-8.9}$ and a slope slightly above $-1.33$), we then studied the GW emissions of points at the same instant $\tau = t^{\rm (c)}$ but slightly off along the string, that is, for $\sigma = \sigma^{\rm (c)} \pm \delta \sigma$, as shown in Fig.~\ref{fig:SlopeCuspVicinitySig}, as well as the GW emissions of the point $\sigma = \sigma^{\rm (c)}$ but at instants slightly before or after the cusp event, that is, for $\tau = t^{\rm (c)} \pm \delta \tau$, as shown in Fig.~\ref{fig:SlopeCuspVicinityTau}. The red curve represents the approach of these points to the cusp ($\sigma < \sigma^{\rm (c)}$ or $\tau < t^{\rm (c)}$) while the black one yields the decline ($\sigma > \sigma^{\rm (c)}$ or $\tau > t^{\rm (c)}$). Note that because we selected space and time intervals of fixed length, the velocity range is not explored with as much detail as one could desire. Still, one can see that the neighbourhood of the cusp is indeed made of highly relativistic points whose GW emissions are cusp-like, namely have a linear high frequency power spectrum behaviour with a slope around $\sfrac{-4}{3}$. This confirms the tendency for such points in the vicinity of cusps to provide the same output as cusps and independent pseudocusps.

\clearpage

\begin{acknowledgments}
The work of M.S. is partially supported by the STFC (UK) under the research grant
ST/L000326/1, and by Perimeter
Institute for Theoretical Physics. Research at Perimeter
Institute is supported by the Government of Canada
through the Department of Innovation, Science and
Economic Development and by the Province of Ontario
through the Ministry of Research and Innovation. The work of M.J.S. is supported by funding from the UK Science and Technology Facilities Council (STFC). The work of T.E. is supported by a KCL GTA studentship.
\end{acknowledgments}


\end{document}